%% file: sample-sigconf.tex
\documentclass[sigconf,nonacm]{acmart}
\AtBeginDocument{%
  }

\setcopyright{acmlicensed}
\copyrightyear{2026}
\acmYear{2026}
\acmDOI{XXXXXXX.XXXXXXX}
\acmConference[KDD '26]{the 32th ACM SIGKDD Conference on Knowledge Discovery and Data Mining}{August 9 to 13,
  2026}{Jeju, Korea}

\acmISBN{978-1-4503-XXXX-X/2018/06}

\usepackage{color}
\definecolor{purple}{RGB}{160,32,240}
 
\usepackage{fontawesome}
\usepackage{amsmath,amsfonts,bm}
\usepackage{multirow}
\usepackage{makecell}
\usepackage[table]{xcolor}
\usepackage{colortbl}
\usepackage[ruled,linesnumbered]{algorithm2e}
\usepackage{algpseudocode}
\usepackage{threeparttable}
\usepackage{enumitem}
\usepackage{marvosym}

\makeatletter
\renewcommand{\@authornotemark}{%
  \g@addto@macro\@currentauthors{%
    \advance\hfuzz by 5pt\relax
    \textsuperscript{\Letter}}}

\renewcommand{\authornote}[1]{%
  \if@ACM@anonymous\else
    \g@addto@macro\addresses{\@authornotemark}%
    \g@addto@macro\@authornotes{%
      \begingroup
      \renewcommand{\thefootnote}{\Letter}%
      \stepcounter{footnote}\footnotetext{#1}%
      \endgroup}%
  \fi}
\makeatother

\newcommand{\ourname}{{HD-Prot}}

\usepackage{subfiles}

\begin{document}

\title{HD-Prot: A Protein Language Model for Joint Sequence-Structure Modeling with Continuous Structure Tokens}

\author{Yi Zhou}
\affiliation{%
  \institution{The Hong Kong Polytechnic University}
  \city{Hong Kong}
  \country{China}
}
\email{echo-yi.zhou@connect.polyu.hk}

\author{Haohao Qu}
\affiliation{%
  \institution{The Hong Kong Polytechnic University}
  \city{Hong Kong}
  \country{China}
}
\email{haohao.qu@connect.polyu.hk}

\author{Yunqing Liu}
\affiliation{%
  \institution{The Hong Kong Polytechnic University}
  \city{Hong Kong}
  \country{China}
}
\email{yunqing617.liu@connect.polyu.hk}

\author{Shanru Lin}
\affiliation{%
  \institution{The Hong Kong Polytechnic University}
  \city{Hong Kong}
  \country{China}
}
\email{lllam32316@gmail.com}

\author{Le Song}
\affiliation{%
  \institution{Mohamed bin Zayed University of Artificial Intelligence}
  \city{Abu Dhabi}
  \country{United Arab Emirates}
}
\email{le.song@mbzuai.ac.ae}

\author{Wenqi Fan} 
\authornote{Corresponding author.}
\affiliation{%
  \institution{The Hong Kong Polytechnic University}
  \city{Hong Kong}
  \country{China}
}
\email{wenqifan03@gmail.com}

\renewcommand{\shortauthors}{Yi Zhou et al.}

\begin{abstract}
\subfile{sections/0-abstract}
\end{abstract}


\keywords{Protein Language Models, AI for Science, Multimodal Protein Modeling, Protein Sequence-Structure Modeling.}

\maketitle

\subfile{sections/1-introduction}

\subfile{sections/2-preliminaries}

\subfile{sections/3-method}
\subfile{sections/4-experiments}
\subfile{sections/5-discussion}

\begin{acks}
The research described in this paper has been partially supported by the General Research Funds from the Hong Kong Research Grants Council (project No. PolyU 15200023, 15206024, and 15224524), and Internal research funds from Hong Kong Polytechnic University (project no. P0059586, P0042693, P0048625, and P0051361). This work was supported by computational resources provided by The Centre for Large AI Models (CLAIM) of The Hong Kong Polytechnic University.
\end{acks}

\bibliographystyle{ACM-Reference-Format}
\bibliography{sample-base}

\appendix
\subfile{sections/6-appendix}

\end{document}

%% file: sections/0-abstract.tex
Proteins inherently possess a consistent sequence-structure duality. 
The abundance of protein sequence data, which can be readily represented as discrete tokens, has driven fruitful developments in protein language models (pLMs).
A key remaining challenge, however, is how to effectively integrate continuous structural knowledge into pLMs. 
Current methods often discretize protein structures to accommodate the language modeling framework, which inevitably results in the loss of fine-grained information and limits the performance potential of multimodal pLMs. 
In this paper, we argue that such concerns can be circumvented: a sequence-based pLM can be extended to incorporate the structure modality through continuous tokens, i.e., high-fidelity protein structure latents that avoid vector quantization.
Specifically, we propose a hybrid diffusion protein language model, \textbf{\ourname{}}, which embeds a continuous-valued diffusion head atop a discrete pLM, enabling seamless operation with both discrete and continuous tokens for joint sequence-structure modeling.
It captures inter-token dependencies across modalities through a unified absorbing diffusion process, and estimates per-token distributions via categorical prediction for sequences and continuous diffusion for structures.
Extensive empirical results demonstrate that \ourname{} achieves competitive performance in unconditional sequence-structure co-generation, motif-scaffolding, protein structure prediction, and inverse folding tasks. Furthermore, our method can perform on par with state-of-the-art multimodal pLMs, despite being developed under limited computational resources (i.e., less than one-tenth the budget for modality extension fine-tuning).
It highlights the viability of simultaneously estimating categorical and continuous distributions within a unified language model architecture, offering a promising alternative direction for multimodal pLMs.
Our code and data are available at \href{https://github.com/EchoChou990919/hdprot}{\textit{https://github.com/EchoChou990919/hdprot}}.
\begingroup
  \renewcommand{\thefootnote}{}
  \footnotetext{This is the long version of the paper to appear at KDD 2026~\cite{zhou2026hdprot}.}
\endgroup

%% file: sections/1-introduction.tex
\section{Introduction}

\begin{figure}
  \centering
  \includegraphics[width=0.9\linewidth]{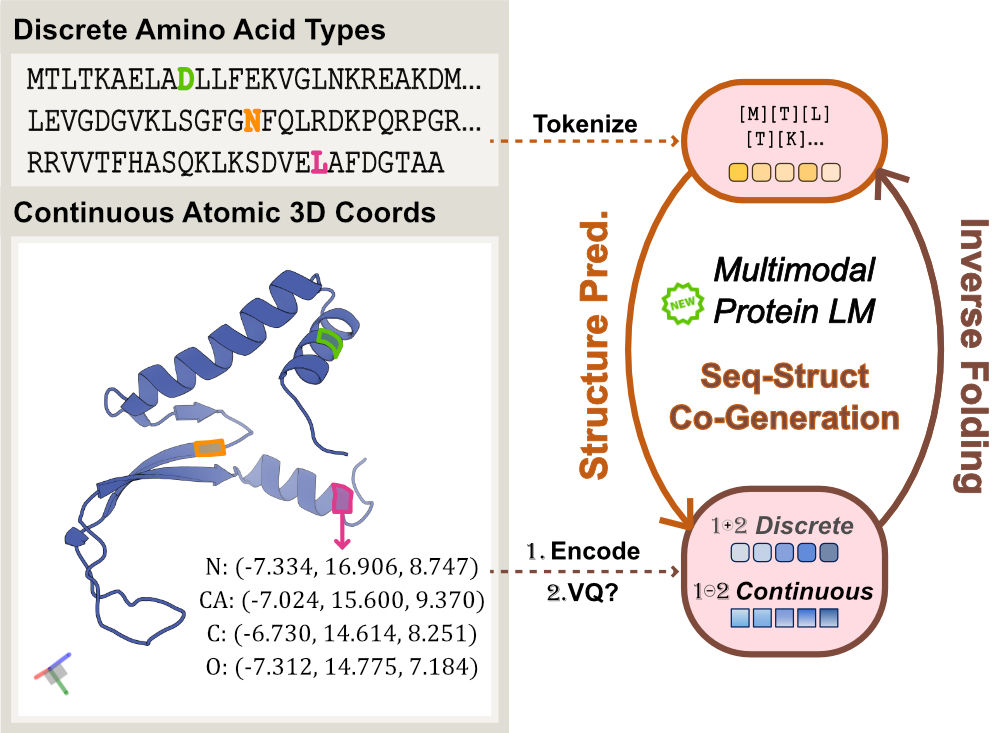}
  \captionsetup{font=small,skip=6pt}
  \caption{
  Background and Motivation. 
  Multimodal pLMs enable joint sequence-structure modeling yet face a fundamental choice in structure representation: discretization or not.
  }
  \label{fig:background}
  \vspace{-0.3cm}
\end{figure}

Proteins, as the fundamental workhorses of life, orchestrate nearly all cellular processes.
Their biological roles are governed by a canonical paradigm~\citep{anfinsen1973principles,liu2026enhancing} -- the \textbf{amino acid sequence} of a protein determines its \textbf{3D structure}, which in turn defines its function. 
As illustrated on the left of Figure~\ref{fig:background}, this relationship highlights both the intrinsic synergy and the distinct nature of protein sequences and structures. 
They are strongly correlated in a biological sense, yet they exhibit significant divergence in data modality: 
the sequence comprises a \textbf{discrete} arrangement of amino acid types, whereas the structure is described by \textbf{continuous}-valued coordinates. 
This duality has motivated an ambitious goal in computational modeling: to develop a unified protein generative model that jointly estimates the distribution of protein sequences and structures.

Benefiting from the greater scale of protein sequence data and the remarkable success of language model pre-training, sequence-first protein language models (\textbf{pLMs}) have established a robust foundation for exploring the vast protein universe~\citep{fan2025computational}. 
Subsequently, an effective path towards the joint modeling of protein sequence and structure is to perform \textbf{modality extension} on pLMs~\citep{hayes2025simulating,wang2024dplm,liu-etal-2025-glprotein}.
These models leverage their strong sequence modeling capability to enable coherent structure learning through a unified semantic space built on shared parameters.
Therefore, as shown on the right of Figure~\ref{fig:background}, multimodal pLMs have the capability to complete complex cross-modal tasks, especially the sequence-structure co-generation, protein structure prediction, and inverse folding.

Nevertheless, as multimodal generative pLMs continue to advance, a critical design choice remains in how to represent protein structure knowledge for language models. To align with standard language model architectures, existing prominent approaches often opt to process the protein structure into discrete tokens. 
Concretely, ESM3~\citep{hayes2025simulating} and DPLM-2~\citep{wang2024dplm} introduce protein structure tokenizers based on quantizers like VQ-VAE~\citep{van2017neural,yu2023language}, thereby representing each structure as a sequence of discrete tokens from learned codebooks. 
However, a fundamental limitation remains: the quantization process inevitably compresses and omits portions of continuous information in pLMs, leading to the loss of fine-grained structural details and imprecise geometric relationships. 
To be specific, this information loss impairs the reconstruction capability of the structure tokenizer first, and ultimately caps the achievable accuracy of structure modeling in multimodal protein language models. 
Having noticed this issue, DPLM-2.1~\citep{hsieh2025elucidating} offers finer-grained structural supervision via bitwise discrete modeling.
While this represents a step toward modeling detailed geometric variations, it still fundamentally relies on discrete representations of 3D structures.

As a promising alternative to discretization approaches, there has been a recent trend toward embracing \emph{continuous tokens} in many multimodal domains, particularly visual-language modeling~\citep{wang2024diffusion,wang2025lavin,li2026molreflect}, with the aim of enhancing continuous information fidelity.
For example, \citet{chen2025softvq} presents an efficient continuous image tokenizer that achieves a high compression ratio while enhancing the semantic richness of the latent space.
\citet{li2024autoregressive} suggests that auto-regressive modeling does not necessarily need to be coupled with discrete and vector-quantized representations. High-quality image generation can be achieved through autoregressive modeling of per-token probability distributions in a continuous-valued space. 
Furthermore, \citet{fan2024fluid} reveals that quantization-based models exhibit slower performance improvements in visual tasks when scaling up model size, compared to models operating on continuous tokens.
In a nutshell, the effectiveness of utilizing continuous tokens has been demonstrated in the visual-language modeling tasks, benefiting from their expressive capability in representing fine-grained knowledge.
Inspired by these cutting-edge advancements, embracing continuous structure tokens alongside natively discrete sequence tokens holds promise for empowering pLMs to achieve high-quality modeling of both protein sequences and structures. 
In this context, a research question arises in this paper:
\textit{Can a protein language model capture the protein structure information in a continuous space, while preserving the extensive knowledge of discrete sequences?}

\begin{figure*}[t]
    \centering
    \includegraphics[width=0.9\linewidth]{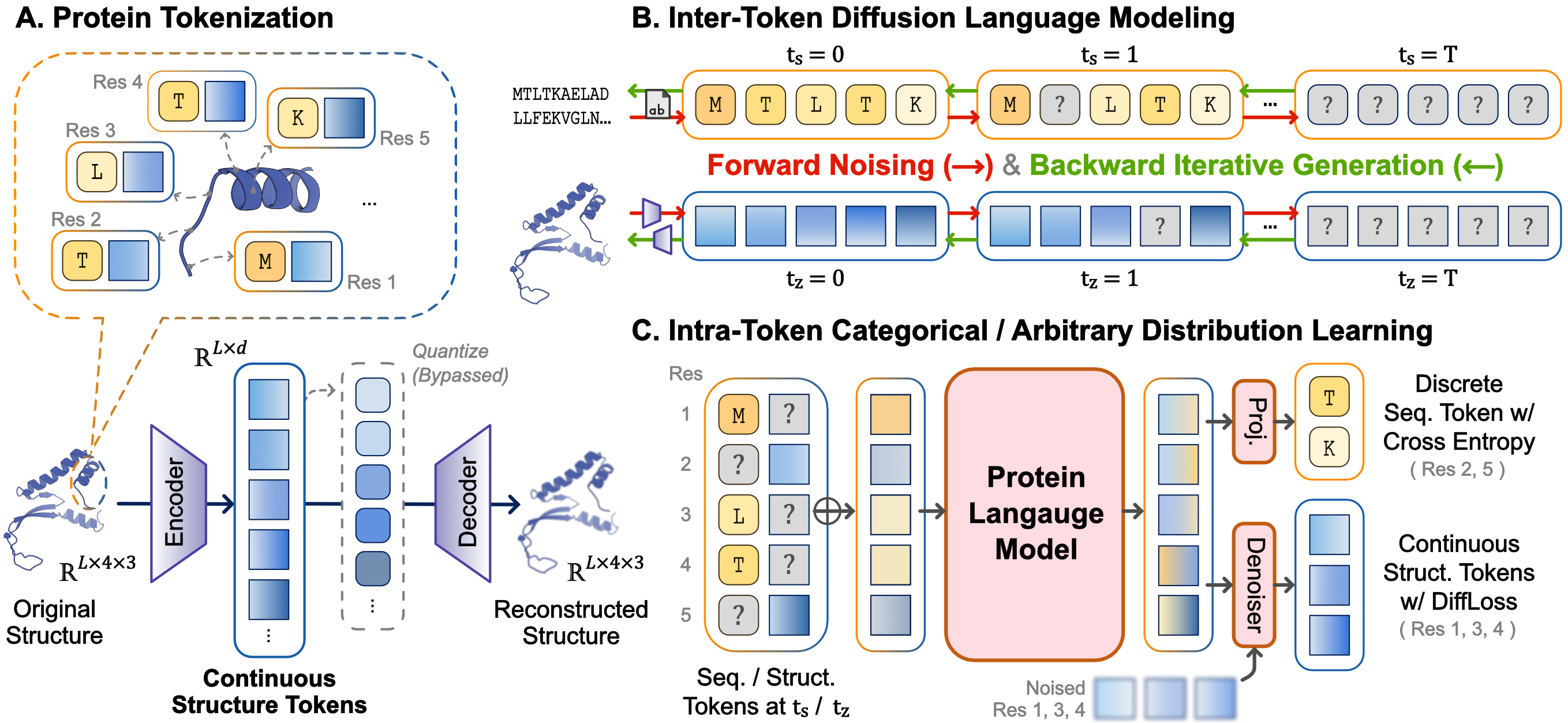}
    \captionsetup{font=small, skip=6pt}
    \caption{
    Overview of Our Proposed \ourname{}. 
    (\textbf{A}) Protein backbone structure is processed into continuous tokens via an advanced non-quantized tokenizer. Each protein is represented by a track of discrete sequence tokens and a track of continuous structure tokens, aligned residue-wise. 
    (\textbf{B}) \ourname{} performs diffusion language modeling to capture inter-token dependencies, wherein both sequence tokens and structure tokens are noised by and denoised from the absorbing state, i.e., the mask. 
    (\textbf{C}) \ourname{} models the protein sequence and structure almost within a unified pLM. 
    Based on the hidden states produced by the pLM, we introduce a categorical head for discrete sequence modeling and a denoising head for continuous-valued structure generation.
    }
    \vspace{-0.3cm}
    \label{fig:overview}
\end{figure*}

In this study, we conclusively address this question with an affirmative answer.
To be specific, we propose \textbf{\ourname{}}, a novel \textbf{H}ybrid \textbf{D}iffusion framework that extends a sequence-only \textbf{Prot}ein language model into a multimodal pLM by incorporating continuous structure tokens. 
Figure~\ref{fig:overview} presents the overall architecture of the proposed \ourname{}.
First, a non-quantized autoencoder is introduced as the protein structure tokenizer, 
where latent representations that can be highly accurately reconstructed into 3D coordinates are considered as continuous structure tokens. 
Globally, the proposed multimodal pLM places the continuous structure tokens on an equal footing with the discrete sequence tokens.
Diffusion language modeling is applied in parallel to both token tracks, involving a noising process that masks protein sequence and structure tokens, followed by a generation process of iterative mask token predictions. 
More concretely, the protein sequence-structure information is residue-wise integrated and consistently processed by the main body of a protein language model.
The per-token probability distribution is estimated via language modeling in a categorical space for sequence and via diffusion modeling in a continuous space for structural knowledge.

In summary, our main contributions are highlighted as follows:
\begin{itemize}[leftmargin=*]
\item This paper highlights the promising paradigm (i.e., \textbf{continuous tokens}) to represent protein structure information within protein language models (pLMs).
We demonstrate that it is effective and efficient to develop a multimodal generative pLM with a non-quantized structure tokenizer and a publicly available sequence-only pre-trained pLM. 

\item We propose \textbf{\ourname{}}, a novel hybrid diffusion framework that bridges the discrete-continuous modality gap in multimodal protein modeling.
In addition to the unified absorbing diffusion language modeling at the inter-token level, the key lies in differentiating the learning of protein sequence and structure at the intra-token level.
Alongside the categorical mask prediction performed on discrete sequence tokens, our model estimates the probability distribution of continuous structure tokens via a diffusion procedure operating on a continuous-valued domain.
\item We conduct comprehensive experiments on four tasks: unconditional sequence-structure co-generation, motif-scaffolding, protein structure prediction, and inverse folding. 
Our proposed \ourname{} models are competitive with representative multimodal pLMs, demonstrating a strong ability to estimate the joint distribution of protein sequence and structure. 
Furthermore, our study provides several valuable insights into practical implementation, specifically regarding robust modality expansion, classifier-free guidance for continuous structure tokens, and ultra-low-cost development that requires \textbf{less than one-tenth budget for modality extension fine-tuning} compared to the representative method DPLM-2.
\end{itemize}

%% file: sections/2-preliminaries.tex
\section{Preliminaries}

\noindent \textbf{Multimodal Protein Modeling.}
A protein can be characterized through its sequence and structure.
For a protein with $ L $ residues, its sequence is defined as $ \bm{s} = \left( s_1, s_2, \dots, s_L \right) $, where each $ s_i \left( 1 \leq i \leq L \right) $ is a categorical variable denotes the amino acid identity of the $i$-th residue, generally involved in 20 standard amino acids $ \mathbb{S}^{20} = \{\texttt{A}, \texttt{R}, \dots, \texttt{V}\} $.
Meanwhile, the protein structure is represented as $ \bm{x} = \left( x_1, x_2, \dots, x_L \right) $, where $ x_i \in \mathbb{R}^{n_i \times 3} $ encoding the Cartesian coordinates of all atoms in the $i$-th residue.
We specifically consider backbone atoms $\{\text{N}, \text{C}_\alpha, \text{C}, \text{O}\}$ that captures the essential structural scaffold, thus simplifying each $ x_i $ to a real-value matrix in $ \mathbb{R}^{4 \times 3} $.

Generative modeling estimates the probabilistic distribution of protein data via a neural network $ \theta $. 
It's expected that a multimodal protein model can holistically understand and explore the protein universe, estimating the joint sequence-structure distribution natively, expressed formally as:
\begin{align}
    p_\theta (\text{Protein}) = p_\theta (\bm{s}, \bm{x}) = p_\theta (s_1, s_2, \dots, s_L, \ x_1, x_2, \dots, x_L).
\end{align}
Whereupon, we are able to perform protein sequence-structure co-generation straightforwardly, and conduct conditional generations across modalities~\citep{wang2025toward,campbell2024generative,wang2024dplm}. 
Such an all-in-one modeling framework is opening up a new direction beyond the cascaded calls of independent sequence/structure generation~\citep{wang2024diffusion,watson2023novo,geffner2025proteina}, structure prediction~\citep{jumper2021highly,lin2023evolutionary}, 
and inverse-folding~\citep{dauparas2022robust,hsu2022learning} models. 

\vspace{0.5em}
\noindent \textbf{Diffusion Language Models.} 
Diffusion models~\citep{ho2020denoising,karras2022elucidating,song2021scorebased} learn to synthesize data by gradually denoising random noise through an iterative process that reverses a predefined noise-adding Markov chain. 
Significant breakthroughs first emerged in the image domain, where diffusion models learn to estimate arbitrary continuous-valued data distributions through iterative denoising of Gaussian noise.
Recent advances have extended diffusion models to language modeling~\citep{deepmind2025gemini,nie2025large,yu2025discrete}, achieving strong performance across a range of benchmarks.
When adopting the mask token $\texttt{<Mask>}$ as an \emph{absorbing} state, diffusion language models operating on categorical distributions retain the basic idea of diffusion models. 
Here we illustrate it following the formulations of~\citet{wang2024diffusion}. 

The \textbf{\textit{forward process}} progressively corrupts an input sentence $ \bm{s}^{(0)} $ over $ T $ diffusion steps through iterative token masking, ultimately transforming all tokens into the mask token.
The $t$-step marginal distribution admits: 
\begin{equation}
    q(\bm{s}^{(t)}|\bm{s}^{(0)}) = \text{Cat} \left( \bm{s}^{(t)}; \bar{\alpha}_t \bm{s}^{(0)} + (1-\bar{\alpha}_t) \bm{q}_{\text{noise}} \right),
\end{equation}
where $\bm{q}_{\text{noise}}$ is a fixed probability vector concentrated on the mask token, and $ \bar{\alpha}_t $ represents the preservation rate of original tokens determined by a masking schedule, satisfying $ \bar{\alpha}_t \to 0$ as $t \to T$. 
The \textbf{\textit{reverse process}} is learned by parameterizing the denoising transition steps:
\begin{equation}
    p_\theta( \bm{s}^{(t-1)} | \bm{s}^{(t)} ) = 
    \textstyle\sum_{\hat{\bm{s}}^{(0)}} q( \bm{s}^{(t-1)} | \bm{s}^{(t)}, \hat{\bm{s}}^{(0)} ) p_\theta ( \hat{\bm{s}}^{(0)} | \bm{s}^{(t)} ),
\end{equation}
where $ \hat{\bm{s}}^{(0)} $ denotes the model's prediction of the full sentence, and transition kernel $ q( \bm{s}^{(t-1)} | \bm{s}^{(t)}, \hat{\bm{s}}^{(0)} ) $ samples a less noisy $ \bm{s}^{(t-1)} $ based on the $ \bm{s}^{(t)} $ and $ \hat{\bm{s}}^{(0)} $.
As simplified by~\citet{austin2021structured}, the \textbf{\textit{training}} undergoes a reweighted masked language modeling: 
\begin{equation}
    \mathcal{L} = \mathbb{E}_{\bm{s}^{(0)}} \left[ \lambda^{(t)} \sum_{i=1}^{L} \bm{1}_{\bm{s}_i^{(t)}=\texttt{<Mask>}} \log p_\theta(\bm{s}_i^{(0)} | \bm{s}^{(t)}) \right],
\end{equation}
where $ L $ represents the length of the corpus and $ \lambda^{(t)} $ is a reweighting term induced from specific masking schedules. 
Eventually, \textbf{\textit{generation}} begins with a sequence of $ \texttt{<Mask>} $ of a specified length, and progressively approaches the realistic sequence $ \bm{s}^{(0)} $ by iterative mask token prediction and remasking that selectively adopts a subset of predicted tokens at each step.

%% file: sections/3-method.tex
\section{The Proposed Method: HD-Prot}
\label{sec:method}

Figure~\ref{fig:overview} provides an overview of \ourname{}. Firstly, there is a protein structure tokenizer capable of transforming between the 3D coordinates and the latent representations, i.e., continuous protein structure tokens. Subsequently, our  \ourname{} framework extends a sequence-only protein language model into a multimodal model by integrating the additional continuous structure tokens. 

\subsection{Continuous Protein Structure Tokens}

As shown in Figure~\ref{fig:overview}.A, a protein structure and its continuous structure tokens are interconverted by a tokenizer.
It basically operates like a non-quantized protein autoencoder, following the encoding-decoding process:
\begin{equation}
\bm{x} \xrightarrow{\text{encoder}} \bm{z} \xrightarrow{\text{decoder}} \hat{\bm{x}},
\end{equation}
where $ \bm{x} \in \mathbb{R}^{L \times 4 \times 3} $ is the input backbone structure, $ \bm{z} \in \mathbb{R}^{L \times d_{\text{struct}}} $
is the continuous protein structure tokens, and $ \hat{\bm{x}} \in \mathbb{R}^{L \times 4 \times 3} $ is the reconstructed 3D coordinates. 
As the foundation for subsequent language modeling,
an ideal tokenizer should learn the structural \textit{equivalence} and \textit{contextual locality} for proteins~\citep{hayes2025simulating}.
Equivalence ensures the protein structure tokens $\bm{z}$ are invariant to the global rotation/translation of $\bm{x}$, enabling the use of a standard, non-equivariant transformer in pLMs.
Contextual locality means each $\bm{z}_i$ (for $1 \leq i \leq L$) primarily corresponds to the local structural environment of residue $i$, ensuring that masking it forces the pLM to learn effective context rather than exploiting global shortcuts. 
To satisfy these requirements, an advanced protein structure autoencoder, salad~\citep{jendrusch2025efficient}, featuring a sparse invariant-point attention (IPA) architecture, is introduced as our protein structure tokenizer. 
Specifically, its latent dimension $d_{\text{struct}} = 20$.

The primary motivation for introducing continuous structure tokens is to minimize information loss in protein structure representation, as evidenced by high-quality protein structure reconstruction. 
As shown in Table~\ref{tab1}, the salad tokenizer outperforms DPLM-2~\citep{wang2024dplm}\footnote{DPLM-2.1 and DPLM-2 share the same protein structure tokenizer based on lookup-free quantization, with a vocabulary size of $2^{13} = 8192$. DPLM-2.1's bitwise discrete modeling provides finer-grained supervision over the 13 binary bits of each token, rather than the corresponding 8192-way index.} and ESM3~\citep{hayes2025simulating} tokenizers on the CAMEO 2022 test set, while also significantly surpassing its VQ-version counterpart~\citep{jendrusch2025efficient}, demonstrating the advantage of avoiding quantization. 
The near-perfect reconstruction capability of the salad tokenizer indicates that its resulting continuous tokens retain virtually all essential protein structure information. 
See Appendix \ref{asec:tokenizer} for more detailed analysis.

\begin{table}[t]\footnotesize
\centering
\captionsetup{font=small, skip=3pt}
\caption{Structure Reconstruction Quality}
\begin{tabular}{cccc}
\toprule
& & \multicolumn{2}{c}{CAMEO} \\
\cmidrule{3-4}
Tokenizer & \#Vocab.
& scRMSD $ \downarrow $ & scTM $ \uparrow $ \\ 
\midrule
DPLM-2 & 8092 & 1.971 ± 1.568 & 0.940 ± 0.071 \\
ESM3 & 4096 & 0.725	± 1.259 & 0.990 ± 0.025 \\
\midrule
salad-vq & 4096 & 1.120 ± 2.025 & 0.979 ± 0.036 \\
\textbf{salad} & - & 0.367 ± 0.803 & 0.997 ± 0.011 \\
\bottomrule
\end{tabular}
\label{tab1}
\end{table}

\subsection{Hybrid Diffusion Protein Language Model}

We propose a hybrid diffusion framework for multimodal protein modeling, which enables a pLM to jointly model a track of discrete sequence tokens $ \bm{s} = \left( s_1, s_2, \dots, s_L \right) $ and a track of continuous structure tokens $ \bm{z} = \left( z_1, z_2, \dots, z_L \right) $.
The common per-residue tokenization allows for a unified 
absorbing diffusion language modeling at the inter-token level, while the discrete/continuous distinction requires separate estimation of categorical/arbitrary distributions at the intra-token level.

\subsubsection{\textbf{Inter-Token Diffusion Language Modeling.}} 

Protein sequences and structures embody a wealth of evolutionary, functional, and folding knowledge, reflected in the relationships between sequence tokens, between structure tokens, and across modalities.
\ourname{} perform unified diffusion language modeling to learn this rich protein knowledge, simultaneously capturing bidirectional contextual dependencies within each modality and cross-modal alignments.

Fundamentally, \ourname{} introduces absorbing states via dedicated mask tokens: $ \bm{m}_{s} $ for sequences and $ \bm{m}_{z} $ structures. 
It configures decoupled schedulers $ t_s\in \{ 0, 1, \dots, T \} $ and $ t_z\in \{ 0, 1, \dots, T \} $ for the protein sequence and structure, respectively. 
Distinct configurations of the two schedulers drive diverse protein modeling tasks, which are detailed in the Appendix~\ref{asec:schedulers}.
As illustrated in Figure~\ref{fig:overview}.B, the \textbf{\textit{forward process}} graudally noise the initial sequence and structure tokens $ ( \bm{s}^{(0)}, \bm{z}^{(0)} ) $ into masks via limited diffusion steps. States at the combined $ (t_s,t_z) $ step is formally defined as:
\begin{equation}
\begin{split}
    q(\bm{s}^{(t_s)} \!\mid\! \bm{s}^{(0)}) &= (1-\frac{t_s}{T}) \bm{s}^{(0)} + \frac{t_s}{T} \bm{m}_s, \\ 
    q(\bm{z}^{(t_z)} \!\mid\! \bm{z}^{(0)}) &= (1-\frac{t_z}{T}) \bm{z}^{(0)} + \frac{t_z}{T} \bm{m}_z. 
\end{split}
\end{equation}
For the sequence track, $(\frac{t_s}{T})L$ randomly selected tokens are replaced with mask token $ \bm{m}_s $ and the remaining $ (1-\frac{t_s}{T})L $ tokens are preserved from the original $ \bm{s}^{(0)}$; so as for the structure track. 
Given that, the model learns to denoise from the fully masked state $ ( \bm{s}^T, \bm{z}^T ) $ through a parameterized \textbf{\textit{reverse process}}: 
\begin{equation}
\begin{split}
    p_\theta( \bm{s}^{(t_s-1)} \!\mid\! \bm{s}^{(t_s)}, \bm{z}^{(t_z)} ) = \textstyle\sum_{\hat{\bm{s}}^{(0)}} &q( \bm{s}^{(t_s-1)} \!\mid\! \bm{s}^{(t_s)}, \hat{\bm{s}}^{(0)} )\\ &p_\theta ( \hat{\bm{s}}^{(0)} \!\mid\! \bm{s}^{(t_s)}, \bm{z}^{(t_z)} ), \\
    p_\theta( \bm{z}^{(t_z-1)} \!\mid\! \bm{z}^{(t_z)}, \bm{s}^{(t_s)} ) = \textstyle\sum_{\hat{\bm{z}}^{(0)}} &q( \bm{z}^{(t_z-1)} \!\mid\! \bm{z}^{(t_z)}, \hat{\bm{z}}^{(0)} )\\ &p_\theta ( \hat{\bm{z}}^{(0)} \!\mid\! \bm{z}^{(t_z)}, \bm{s}^{(t_s)} ).
\end{split}
\label{reverse}
\end{equation}
For the sequence track, $ \hat{\bm{s}}^{(0)} $ denotes the model's prediction of the initial state based on the partially masked states at $ ( t_s, t_z ) $, and the less noisy $ \bm{s}^{(t_s-1)} $ is sampled conditioned on the $ ( \bm{s}^{(t_s)}, \bm{z}^{(t_z)} ) $ and $ \hat{\bm{s}}^{(0)} $ via the transition kernel $ q $; so as for the structure track.

\subsubsection{\textbf{Intra-Token Categorical / Arbitrary Distribution Learning.}}

To accommodate the distinct characteristics of multimodal protein data, we introduce two intra-token learning channels: categorical prediction for protein sequence tokens and continuous-valued estimation for the structure tokens.
As shown in Figure~\ref{fig:overview}.C,  the partially masked sequence and structure tokens are fused at the input and processed through a protein language model (pLM): 
\begin{equation}
\begin{split}
    \bm{c} &= \text{pLM}\left( \bm{c}_{\text{seq}} + \bm{c}_{\text{struct}} \right), \\ 
    \bm{c}_{\text{seq}} = \text{embed}&\left( \bm{s}^{(t_s)} \right),
    \quad
    \bm{c}_{\text{struct}} = \text{norm}\left( \bm{z}^{(t_s)} \right)W_{\text{in}},
\end{split}
\end{equation}
where the sequence tokens $ \bm{s}^{{(t_s)}} $ are mapped to embeddings $ \bm{c}_{\text{seq}} \in \mathbb{R}^{L \times d_{\text{hidden}}} $ via the pLM's embedding module, and the structure tokens $ \bm{s}^{{(t_z)}} $ are transformed to $ \bm{c} \in \mathbb{R}^{L \times d_{\text{hidden}}} $ via a layer normalization and a linear projection $ W_{\text{in}} \in \mathbb{R}^{d_{\text{struct}} \times d_{\text{hidden}}} $.
A pLM receives the fused sequence-structure representation and produces the deeply integrated protein representation $ \bm{c} $.
Together, the element-wise summation operation and the shared language model position encoding guarantee \textit{residue-by-residue sequence-structure alignment}~\citep{hayes2025simulating}.

Subsequently, the model learns to estimate the \textit{per-token distribution} through a reweighted cross-entropy loss and diffusion loss~\citep{li2024autoregressive} for sequence and structure tokens, respectively: 
\begin{equation}
\begin{split}
    \mathcal{L}_{\text{seq}} = \mathbb{E}_{\bm{s}^{(0)}} &\left[ \lambda^{(t_s)}_{\text{seq}} 
    \sum_{i=1}^{L} \bm{1}_{\bm{s}_i^{(t_s)}=\bm{m}_s} 
    \log p(\bm{s}_i^{(0)} | \bm{c}_i) \right],
    \\
    p(\bm{s}_i^{(0)} \!\mid\! \bm{c}_i&) = \text{Softmax}\left( \text{Projector}\left( \bm{c}_i \right) \right);
\end{split}
\end{equation}
\begin{equation}
\begin{split}
    \mathcal{L}_{\text{struct}} = \mathbb{E}_{\bm{z}^{(0)}} &\left[ \lambda^{(t_z)}_{\text{struct}} 
    \sum_{i=1}^{L} \bm{1}_{\bm{z}_i^{(t_z)}=\bm{m}_z} 
    \Vert \epsilon - \hat{\epsilon}_i \Vert^2 \right],
    \\
    \hat{\epsilon}_i = \text{Denoiser}&\left( \sqrt{\bar{\alpha}_{t'}} \bm{z}_i^{(0)} + \sqrt{1-\bar{\alpha}_{t'}} \epsilon, t', \bm{c}_i \right);
\end{split}
\end{equation}
where the \text{Projector} predicts the categorical logits over the vocabulary of protein sequence tokens, while the \text{Denoiser} is a noise predictor under the typical DDPM framework~\citep{ho2020denoising}.
$ \lambda^{(t_s)} $ and $ \lambda^{(t_z)} $ are reweighting coefficients that control the trade-off between micro and macro perceptions during protein sequence and structure modeling.
For residue $ i $ with the ground-truth structure token $ \bm{z}_i^{(0)} $, the \text{Denoiser} learns to estimate a Gaussian noise 
$ \epsilon \in \mathbb{R}^{d_{\text{struct}}} $ 
based on three factors: the residue representation $ \bm{c}_i $ containing its 
contextual information; a timestamp $ t' $ randomly sampled from $ \{ 1, 2, \dots, T' \} $; and a noised token at the $ t' $ step, formulated as $ \sqrt{\bar{\alpha}_{t'}} \bm{z}_i^{(0)} \!+\! \sqrt{1\!-\!\bar{\alpha}_{t'}} \epsilon $, where the $ \bar{\alpha}_{t'} $ is defined by a noise scheduler~\citep{ho2020denoising,nichol2021improved}. 

Eventually, all learnable parameters are optimized through: 
\begin{equation}
    \mathcal{L} = \mathcal{L}_{\text{struct}} + \gamma \mathcal{L}_{\text{seq}},
\end{equation}
where $\gamma$ balances the focus between protein sequence and structure modeling. 
Detailed settings of training hyperparameters are explained in \ref{asec:training}.

\subsubsection{\textbf{Multimodal Protein Generation.}}
With the \textit{per-token distribution} learned in parallel, the sequence and structure tracks of the pLM employ different samplers, correspondingly.
Taking residue $ i $ with temporary condition representation $ \bm{c}_i^{(t_s)} $ as an example, the {masked sequence prediction} can be done by the categorical sampler:
\begin{equation}
    p(\hat{\bm{s}}_i^{(0)} | \bm{s}^{(t_s)}, \bm{z}^{(t_s)}) = \text{Softmax}\left( \text{Projector}\left( \bm{c}_i^{(t_s)} \right) / \tau_s \right),
\end{equation}
where $ \tau_s $ is the generation temperature for protein sequences.
Meanwhile, given a hidden condition representation $ \bm{c}_i^{(t_z)} $, the {masked structure prediction} undergoes a reverse diffusion procedure of DDPM~\citep{ho2020denoising}, generating $ \hat{\bm{z}}_i^{(0)} $ from a Gaussian noise over $ T' $ steps: 
\begin{equation}
    \hat{\bm{z}}^{(t'-1)}_i = \frac{1}{\sqrt{\alpha_{t'}}} \left( \hat{\bm{z}}^{(t')} - \frac{1-\alpha_{t'}}{\sqrt{1-\bar{\alpha}_{t'}}} \text{Denoiser}(\hat{\bm{z}}_i^{(t')}, t', \bm{c}_i^{(t_z)}) \right) + (\sigma_{t'} \delta)\tau_z,
\end{equation}
where $ \tau_z $ controls the generation temperature for protein structure, $\delta$ is randomly sampled from the Gaussian distribution $\mathcal{N}(\mathbf{0}, \bm{I})$, and $\sigma_{t'}$ represents the noise level at denoising step $t'$~\citep{li2024autoregressive}. 
The reverse procedure of DDPM naturally supports classifier-free guidance (CFG)~\citep{ho2022classifier}. In the context of multimodal protein modeling, we can consider the whole sequence track as the guidance condition, aiming to generate more self-consistent protein structure tokens.

To recap, the \textbf{\textit{multimodal protein generation}}
process follows the reverse diffusion language modeling process formulated in Equation~\ref{reverse}, starting from a state where the protein sequence and structure are either fully masked (for unconditional sequence-structure co-generation) or partially masked (for motif-scaffolding, structure prediction, and inverse folding). 
For each step of the iterative generation, the model predicts all masked tokens, then selectively retains a certain proportion of these predictions while re-masking the remainder for the next step. 
Detailed generation procedures are also provided in the Appendix~\ref{asec:cogen}-\ref{asec:cfg}.

%% file: sections/4-experiments.tex
\section{Experiments}
\label{sec:experiments}

We primarily evaluate \ourname{} models on four foundational tasks: unconditional protein sequence-structure co-generation (Section~\ref{sec4.1}), motif-scaffolding (Section~\ref{sec4.2}), protein structure prediction (Section~\ref{sec4.3}), and inverse folding (Section~\ref{sec4.4}).
Please refer to the Appendix~\ref{app:imp_details} for implementation details, including the training dataset and training process of our model, the implementation of baseline models, and the clarification of evaluation metrics. 

\subsection{Unconditional Protein Sequence-Structure Co-Generation} \label{sec4.1}

In this task, models are required to generate protein sequences and structures simultaneously, using only the specified protein length as input.
We compare our \ourname{} model with three state-of-the-art protein co-generation methods, i.e., MultiFlow~\citep{campbell2024generative}, La-Proteina~\citep{geffner2025laproteina} and PLAID~\citep{lu2025all}, and three multimodal pLMs, i.e., ESM3~\citep{hayes2025simulating}, DPLM-2~\citep{wang2024dplm}, and DPLM-2.1~\citep{hsieh2025elucidating}.
For protein lengths of 100, 200, 300, 400, and 500, we generate 100 proteins per method at each length, with five independent runs using different random seeds.
Moreover, $ 5 \times 100 $ distinct PDB proteins are randomly selected to serve as reference samples.

\subsubsection{\textbf{Quantitative Analysis.}}
Referring to \citet{campbell2024generative} and \citet{wang2024dplm}, the generation results are quantitatively evaluated by three sets of metrics, namely the designability, diversity, and novelty. 
(1) \textbf{Designability}.
A generated protein is considered designable if its sequence is foldable and its structure is consistent with the sequence's structure prediction result.
The \emph{foldability} of a sequence is assessed using the \texttt{pLDDT} score given by ESMFold~\citep{lin2023evolutionary} during structure prediction.
Meanwhile, \emph{self-consistency} between the co-generated structure and the ESMFold-predicted structure is evaluated using backbone \texttt{scRMSD} and \texttt{scTM}. 
(2) \textbf{Diversity}.
For a set of generated proteins, we calculate the number of clusters derived by Foldseek~\citep{van2024fast} with the TM-score threshold at $ 0.5 $ and $ 0.95 $, resulting in \texttt{\#Cluster@50} and \texttt{\#Cluster@95}.
(3) \textbf{Novelty}.
A generated protein is novel if it is dissimilar to well-known proteins, e.g., the PDB~\citep{wwpdb2019protein} or AlphaFoldDB-SwissProt~\citep{jumper2021highly} proteins.
We search for the most similar protein in a reference database and record the TM-score values, leading to \texttt{pdb-TM} and \texttt{sp-TM}. 

\begin{table*}[t]\footnotesize
\centering
\begin{threeparttable}
\captionsetup{font=small, skip=3pt}
\caption{Evaluation of Unconditional Protein Sequence-Structure Co-Generation. 
}
\begin{tabular}{lccccccc}
\toprule
\multirow{2}{*}[-1ex]{\makecell[l]{Models ($\#$Params, \\$\#$Training Sample)}} & \multicolumn{3}{c}{Designability} & \multicolumn{2}{c}{Diversity} & \multicolumn{2}{c}{Novelty} \\
\cmidrule(l){2-4} \cmidrule(l){5-6} \cmidrule(l){7-8} 
& 
pLDDT $\uparrow$ & scRMSD $\downarrow$ & scTM $\uparrow$ & \#Cluster@50 $\uparrow$ & \#Cluster@95 $\uparrow$ & pdb-TM $\downarrow$ & sp-TM $\downarrow$ \\ 
\midrule
MultiFlow (21M{, 22.8K}) & {79.271 ± 7.978} & {2.955 ± 4.252} & {0.937 ± 0.100} & 
{55.12 ± 15.79} & {100.00 ± 0.00} & 0.828 ± 0.054 & 0.826 ± 0.063 \\
* MultiFlow & 61.519 & 9.306 {± 8.499} & 0.750 {± 0.163} & 
49.00 & - & - & - \\
{La-Proteina (158M, 550K)} & {80.152 ± 10.51} & {4.477 ± 6.652} & {0.923 ± 0.141} & 
{64.32 ± 9.586} & {100.00 ± 0.00} & 0.801 ± 0.087 & 0.786 ± 0.085 \\
La-Proteina$_{\text{ tri}}$ (167M, 550K) & {83.770 ± 10.13} & {3.260 ± 6.317} & {0.953 ± 0.119} & {40.60 ± 22.45} & {100.00 ± 0.00} & 0.839 ± 0.092 & 0.818 ± 0.088 \\
PLAID (100M, 57M) & 52.682 ± 18.78 & 9.888 ± 9.144 & 0.764 ± 0.220 & 56.35 ± 9.443 & 99.54 ± 1.046 & 0.816 ± 0.107 & 0.823 ± 0.103 \\
PLAID (2B, 57M) & 57.485 ± 21.87 & 9.008 ± 9.214 & 0.787 ± 0.230 & 61.27 ± 8.993 & 95.12 ± 6.244 & 0.822 ± 0.119 & 0.834 ± 0.113 \\
ESM3 (1.4B{, 1.08B}) & {76.079 ± 13.53} & {31.98 ± 33.87} & {0.762 ± 0.221} & 
{48.00 ± 16.82} & {96.24 ± 7.704} & 0.873 ± 0.104 & 0.899 ± 0.077 \\
\midrule
\rowcolor{gray!10} DPLM-2 (150M{, 220K}) & {82.525 ± 7.754} & {5.125 ± 5.101} & {0.895 ± 0.112} & 
{43.28 ± 7.871} & {83.08 ± 8.665} & 0.920 ± 0.058 & 0.932 ± 0.055 \\
DPLM-2 (650M{, 220K}) & {81.920 ± 8.643} & {4.899 ± 5.523} & {0.906 ± 0.105} & 
{52.40 ± 6.083} & {82.40 ± 8.765} & 0.921 ± 0.068 & 0.934 ± 0.066 \\
DPLM-2.1 (650M{, -}) & {84.773 ± 7.719} & {5.076 ± 5.155} & {0.898 ± 0.114} & 
{60.40 ± 5.766} & {89.28 ± 6.059} & 0.900 ± 0.095 & 0.930 ± 0.064 \\
\midrule
\rowcolor{gray!10} \textbf{\ourname{}} (155M{, 210K}) & {80.646 ± 11.07} & {4.629 ± 4.709} & {0.887 ± 0.127} & 
{44.32 ± 7.409} & {78.32 ± 12.84} & 0.896 ± 0.114 & 0.919 ± 0.102 \\
\textbf{\ourname{}} (670M{, 210K}) & {81.099 ± 9.832} & {4.899 ± 4.534} & {0.878 ± 0.126} & 
{51.16 ± 6.593} & {86.08 ± 4.672} & 0.897 ± 0.107 & 0.917 ± 0.099 \\
\midrule
PDB Proteins & {79.075 ± 13.03} & {4.669 ± 7.683} & {0.905 ± 0.143} & 
{55.80 ± 5.671} & {78.40 ± 3.499} & - & - \\
\bottomrule
\end{tabular}
\label{tab2}
\begin{tablenotes}
\footnotesize
\item * denotes the performance of the MultiFlow variant (w/o data distillation) reported by~\citet{wang2024dplm}.
\end{tablenotes}
\end{threeparttable}
\end{table*}

Table~\ref{tab2} and Appendix Figure~\ref{afig:uncon_cogen} 
shows the overall comparison results.
Notably, natural proteins with absolute self-consistency still do not achieve perfect scores on these metrics, despite their strong overall performance. 
This can be somehow attributed to the use of ESMFold's predicted structure as a reference, which introduces a certain level of model bias.
Therefore, we consider the performance of natural proteins as a special baseline: if a model surpasses this baseline, it may suggest an idealized outcome.
For example, MultiFlow, enhanced with data distillation, substantially outperforms other models as well as the natural protein baseline (i.e., PDB proteins) in designability and diversity. 
However, this may be because the model fits the simplified distribution of the distilled data instead of learning the more complex natural protein knowledge~\citep{campbell2024generative,wang2024dplm}. 
When the data samples distilled by ProteinMPNN are removed, MultiFlow’s performance degrades substantially, particularly collapsing in its sequence generation ability.
Similarly, La-Proteina~\citep{geffner2025laproteina} achieves state-of-the-art performance by being sufficiently scaled up with great computational efforts on a carefully curated set of representative proteins from the AlphaFold database.

In pLMs that generally have more solid sequence modeling ability, ESM3 is significantly lagging behind.
Although ESM3 undergoes extensive pre-training of masked language modeling across dynamic mask rates, it still struggles with prediction under high mask rates, 
resulting in suboptimal performance in unconditional protein sequence-structure co-generation.
In contrast, the DPLM families show a great performance.
The proteins they generate exhibit self-consistency and diversity similar to that of natural proteins, despite the cost of novelty.
More importantly, our \ourname{} models exhibit competitive performance with the DPLM families. \ourname{} (155M) presents a high degree of designability common to that of DPLM-2 (150M); 
Meanwhile, all \ourname{} (670M), DPLM-2 (650M), and DPLM-2.1 (650M) models show a similar trend of enhancing the diversity of generation as the scale of parameters grows. 
Moreover, an interesting observation is that our \ourname{} performs better in the scRMSD compared to the scTM, i.e., excels more in the generation of local structure details. 
This advantage may stem from the use of continuous structure tokens, which capture fine-grained residue-level conformational details more accurately.

Interestingly, PLAID~\citep{lu2025all} leads a trend of latent diffusion modeling based on a continuous joint distribution of protein sequence and structure~\citep{lu2025tokenized,meshchaninov2024diffusion}, sharing a certain degree of conceptual overlap with our work. 
The key distinction is that PLAID constructs a pure continuous diffusion model to learn a latent space between ESM and ESMFold, rather than building a new pLM with deeply fused modalities. 
Evaluation results show that PLAID lacks the sequence modeling capability that pLMs excel at, yielding proteins with lower designability compared to our \ourname{}.

Furthermore, it is worth noting that the \ourname{} framework can be implemented with great computational efficiency. 
Due to the limitations of computational resources, we make many compromises in the implementation, especially the pre-cached tokenization results, mixed-precision training, and a smaller batch size. 
Ultimately, \ourname{} can be successfully trained on only 1$\sim$2 GPUs. 
If converting the $ \text{number} \times \text{device} \times \text{days} $ into the rental price, our modality extension fine-tuning cost is less than \textbf{\emph{one-tenth}} of that of DPLM-2 
(explained in the Appendix~\ref{asec:cost}), 
ensuring a fair comparison as both models are built upon the same foundation model and use similar training data. 
Besides, the inference efficiency of multimodal pLMs is also discussed in the Appendix~\ref{asec:infer_efficiency}.

\subsubsection{\textbf{Qualitative Analysis.}}
\begin{figure*}
    \centering
    \includegraphics[width=\linewidth]{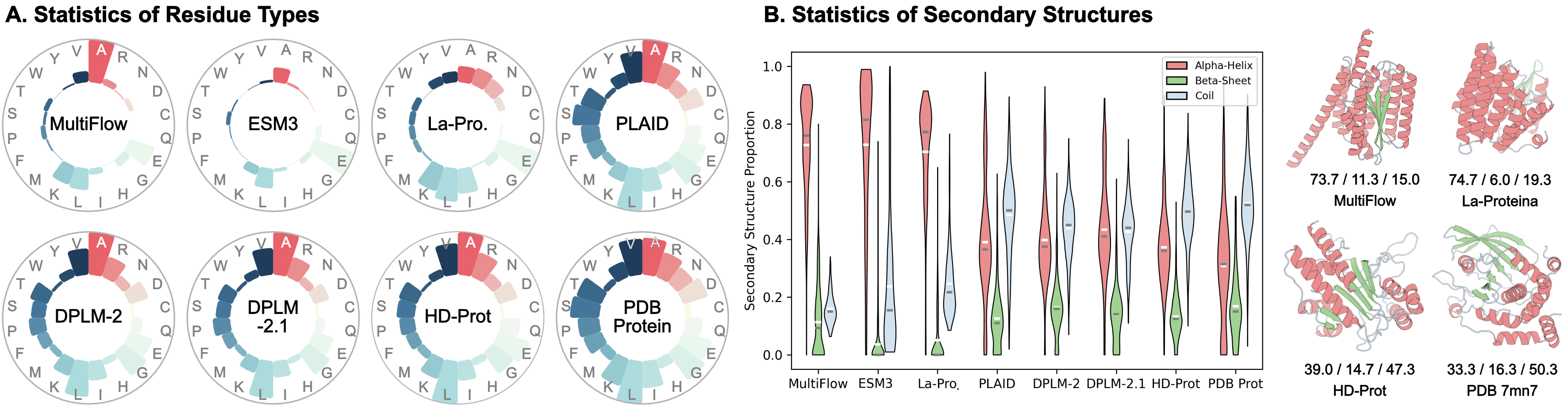}
    \captionsetup{font=small,skip=6pt}
    \caption{Qualitative Analysis. (\textbf{A}-\textbf{B}) The proteins generated by \ourname{} exhibit a similar distribution of residue types and secondary structure proportions compared to native proteins. 
    }
    \label{fig4}
\end{figure*}

Figure~\ref{fig4} shows the assessments of protein samples generated by each method. 
On the sequence modality, we compute the amino acid frequencies and visualize the distributions using Nightingale rose charts in Figure~\ref{fig4}.A. The sequences generated by MultiFlow contain an unusually high proportion of Alanine (\texttt{A}), and the categorical distributions learned by ESM3 and La-Proteina are biased toward Glutamic (\texttt{E}). 
In contrast, PLAID, DPLM-2 series, and \ourname{} models can generate protein sequences with a relatively balanced ratio of various amino acids, similar to natural proteins. 
Nevertheless, the PLAID sequences are hardly foldable (avg. pLDDT $<60$), falling far behind that of DPLM-2, \ourname{}, and native proteins (avg. pLDDT $\approx 80$).
On the structure modality, statistics of the proportion of secondary structures are presented in Figure~\ref{fig4}.B. MultiFlow, ESM3, and La-Proteina exhibit a strong bias toward generating alpha-helices over beta-sheets and coils, whereas PLAID, DPLM-2 series, and \ourname{} produce proteins with secondary structure distributions that more closely resemble natural compositions. 
We select case samples with a length of 300 residues and a secondary structure ratio that is close to the corresponding average values. It is observed that structures generated by MultiFlow usually appear ordered, with one clump after another of alpha-helices or beta-sheets, and very few coils. However, native structures and \ourname{}-generated structures could contain nearly half of them as coils, therefore looking more ``flexible".

We attribute the similar 
performance of DPLM-2 series and our \ourname{} model to the closely aligned training datasets (Appendix \ref{asec:training_dataset}). 
These results indicate that, for building multimodal protein models on top of sequence-based pLMs, quantization-based tokenization of structures is not the only viable approach. Effectively integrating continuous structural representations into pLMs offers an alternative route that also successfully captures the underlying data distribution. 
Besides, case studies of \ourname{} can be found in the Appendix~\ref{asec:mode_analysis}, including visualizations of some excellent co-generation results, and an analysis of the typical failure mode.

\subsubsection{\textbf{Ablation Study.}}
Among various factors related to the implementation of \ourname{}, we identify three key findings. Experimental results are presented in Table~\ref {tab3}. 
First, \emph{the protein sequence foundation model is of great significance}. 
As shown in row 1, when training from scratch, our current data ($\sim$210K proteins) remains insufficient to support effective language modeling, even for pLM at a relatively small scale of 150M parameters.

Second, we need to skillfully control the scale of fine-tuning, achieving \emph{a balance between retaining foundational protein sequence knowledge and acquiring more protein structure information.}
When performing modal extension based on a sequence-only pLM, \ourname{} can encounter mild but non-negligible sequence modal collapse during full-model fine-tuning, particularly for larger models. 
For instance, a 150M-parameter pLM retains high sequence quality after full-model fine-tuning (rows 3, 4), whereas a 650M-parameter pLM, due to its high capacity and our limited training data, suffers from sequence knowledge forgetting (row 7).
Crucially, once the modality collapse of protein sequence is prevented, scaling up through larger foundation models or by expanding learnable parameters (e.g., via LoRA) enables the model to capture a broader data distribution and generate more diverse proteins, as evidenced by rows 2, 4, and 6.

\begin{table}[t]\footnotesize
\vspace{-0.3cm}
\centering
\begin{threeparttable}
\captionsetup{font=small,skip=3pt}
\caption{Ablation Study.}
{
\begin{tabular}{ccccccc}
\toprule
& FM & \#Param (M) & CFG & pLDDT $\uparrow$ & scRMSD $\downarrow$& \#CL@50 $\uparrow$ \\
\midrule
1 & $ \times $ & 150 (150) + 5 & - & 73.100 & 6.798 & - \\
2 & $ \checkmark $ & 32 (150) + 5 & $ \checkmark $ & 81.520 & 4.580 & 39.610 \\
3 & $ \checkmark $ & 150 (150) + 5 & $ \times $ & {80.155} & {4.804} & {42.040} \\
4 & $ \checkmark $ & 150 (150) + 5 & $ \checkmark $ & {80.646} & {4.629} & {44.320} \\
\midrule
5 & $ \checkmark $ & 91 (650) + 20 & $ \times $ & {80.132} & {5.084} & {48.680} \\
6 & $ \checkmark $ & 91 (650) + 20 & $ \checkmark $ & {81.099} & {4.899} & {51.160} \\
7 & $ \checkmark $ & 650 (650) + 20 & - & 73.455 & 6.970 & - \\
\bottomrule
\end{tabular}
}
\label{tab3}
\begin{tablenotes}
\footnotesize
\item \#Param denotes the "tunable (base model) + denoising head" parameters.
\end{tablenotes}
\end{threeparttable}
\vspace{-0.3cm}
\end{table}

Third, \emph{classifier-free guidance (CFG)~\citep{ho2022classifier} can help generate high-quality continuous structure tokens}. 
Indeed, the unconditional sequence-structure co-generation in \ourname{} can be viewed as iterative per-token sampling under cross-modal conditioning. 
When generating a specific protein structure token, replacing the sequence track with masks essentially performs a special ``unconditional" generation. 
Therefore, we can employ the classic classifier-free guidance to steer the sampling of continuous structure tokens towards better consistency by combining the "conditional" and "unconditional" predictions. 
It is observed that CFG improves protein sequence-structure consistency without impairing generation diversity (rows 3-4, 6-7). 
Additionally, Appendix~\ref{asec:temp_cfg_uncon_cogen} provides detailed ablations of the combined effects of two main sampling hyperparameters, i.e., the sampling temperature of structure tokens and the CFG scale.

\subsection{Motif-Conditioned Protein Sequence-Structure Co-Generation}
\label{sec4.2}

\begin{table}[t]\footnotesize
\vspace{-0.3cm}
\centering
\captionsetup{font=small, skip=3pt}
\caption{Motif-Scaffolding Results.
}
{
\begin{tabular}{lcc}
\toprule
& \multirow{2}{*}{\makecell[c]{\#Solved Problems {/ 24}\\mean (min, max)}} & \multirow{2}{*}{\makecell[c]{Success Rate}} \\
\\
\midrule
ESM3 & 19.6 (19, 20) & 30.1\% ± 0.3\%  \\
DPLM-2 (150M) & {15.6 (14, 17)} & {20.0\% ± 7.0\%} \\
DPLM-2 (650M) & {17.8 (16, 19)} & {27.7\% ± 0.8\%} \\
\midrule
\ourname{} (155M) & {18.2 (18, 19)} & {15.9\% ± 0.3\%} \\
\ourname{} (670M) & {19.4 (19, 21)} & {24.1\% ± 1.1\%} \\
\bottomrule
\end{tabular}
}
\label{tab4}
\vspace{-0.3cm}
\end{table}
Motif refers to a significant local pattern within a protein, while scaffold denotes the overall global structural framework that supports these motifs.
Motif-scaffolding aims to design a stable protein scaffold that correctly positions one or more specified motifs.
We adopt the experimental setup of \citet{yim2024improved} and \citet{wang2024dplm} across 24 motif-scaffolding tasks, sampling 100 scaffolds for each task in a run. 
The scaffold length and motif order are determined according to specifications. 
While focusing on the sequence-structure co-generation, both the sequence and structure of the motif are provided as the input condition. 
A motif-scaffolding case is considered successful if it meets the requirements of overall designability and local motif preservation at a time. 
Specifically, the criteria require $\texttt{scTM} > 0.8$ and $\texttt{motif-RMSD} < 1.0$ Å~\citep{wang2024dplm}, ensuring both self-consistency between the predicted structure of the generated sequence and the directly generated structure, as well as accuracy in the predicted motif structure relative to the native motif. 

We evaluate \ourname{} against ESM3 and DPLM-2 based on the number of solved problems and success rate.
Table~\ref{tab4} and Appendix~\ref{asec:scaffolding_problems} summarize the results of five repetitions of sampling with different random seeds, demonstrating that \ourname{} effectively generates protein scaffolds that precisely match the given motifs. 
It solves 19.4 of 24 subtasks on average, outperforming DPLM-2 and competing closely with ESM3.
Additionally, \ourname{} achieves a comparable average success rate, with approximately one-quarter of all $ 24*100 $ generated samples meeting the success criteria.
These results, while preliminary, underscore the potential of protein sequence-structure co-generation as an effective strategy for advancing conditional protein design. 
Besides, an analysis of the sampling hyperparameters of \ourname{} is in Appendix~\ref{asec:temp_cfg_scaffolding}.

\subsection{Protein Structure Prediction}
\label{sec4.3}

\begin{table}[t]\footnotesize
\centering
\captionsetup{font=small,skip=3pt}
\caption{Evaluation of Protein Structure Prediction.
}
{
\setlength{\tabcolsep}{3pt}
\begin{tabular}{lcccc}
\toprule
~ & \multicolumn{2}{c}{CAMEO} & \multicolumn{2}{c}{PDB Date Split} \\
\cmidrule(l){2-3} \cmidrule(l){4-5}
Model & RMSD $ \downarrow $ & TM-score $ \uparrow $ & RMSD $ \downarrow $ & TM-score $ \uparrow $ \\
\midrule
MultiFlow & 17.952 ± 5.991 & 0.500 ± 0.146 & 15.618 ± 4.487 & 0.524 ± 0.136 \\
ESM3 & 5.377 ± 6.303 & 0.860 ± 0.168 & 4.042 ± 4.824 & 0.883 ± 0.150 \\
\midrule
\rowcolor{gray!10} DPLM-2 (150M) & 9.919 ± 6.994 & 0.720 ± 0.189 & 7.833 ± 6.004 & 0.765 ± 0.169 \\
DPLM-2 (650M) & 7.483 ± 6.126 & 0.786 ± 0.170 & 5.253 ± 5.143 & 0.836 ± 0.144 \\
DPLM-2.1 & 6.272 ± 6.202 & 0.824 ± 0.166 & 2.869 ± 3.942 & 0.915 ± 0.113 \\
\midrule
\rowcolor{gray!10} HD-Prot (155M) & 9.185 ± 6.316 & 0.719 ± 0.201 & 6.229 ± 5.391 & 0.781 ± 0.181 \\
HD-Prot (670M) & 7.468 ± 6.004 & 0.769 ± 0.177 & 5.001 ± 4.565 & 0.827 ± 0.153 \\ 
\bottomrule
\end{tabular}
}
\label{tab5}
\vspace{-0.3cm}
\end{table}

Protein structure prediction aims to infer the 3D structure of a protein according to its amino acid sequence~\citep{jumper2021highly,lin2023evolutionary}. 
In the context of joint sequence-structure modeling, protein structure prediction is also considered a sequence-conditioned structure generation task. 
Following the experimental setup of established studies~\cite{wang2024dplm,hsieh2025elucidating}, we evaluate the structure prediction capability of multimodal protein generative models via two datasets, i.e., CAMEO 2022, and a PDB Date Split curated by \citet{campbell2024generative}. 
The structure prediction results are compared to the corresponding native structures, and the \texttt{RMSD} and \texttt{TM-score} are calculated to assess the prediction accuracy. 

Table~\ref{tab5} presents the comparison between \ourname{} and four multimodal protein generative models, where all predictions with randomness are repeated five times with different seeds. 
Firstly, compared with the unconditional protein sequence-structure co-generation results (Table~\ref{tab2}), MultiFlow and ESM3 present totally different capabilities in protein structure prediction.
Due to the reliance on non-natural distillation data, MultiFlow lacks the ability to understand the natural sequence arrangement as well as the sequence-to-structure folding rules. 
Meanwhile, given the complete sequence information, the ultra-large-scale pre-trained ESM3 model can accurately infer the corresponding structural information. 
Notably, \ourname{} performs better or comparable to the DPLM-2 at both $\sim$150M and $\sim$650M scales. During the training of \ourname{}, it has never seen a situation where the sequence track is completely given, and the structure track is fully masked. This absolutely zero-shot protein structure prediction performance indicates that \ourname{} has acquired considerable sequence-structure cross-modal capabilities. 
Besides, the explanation of the sampling hyperparameters of \ourname{} can be found in Appendix~\ref{asec:temp_cfg_folding}.

\subsection{Inverse Folding}
\label{sec4.4}

Inverse folding, also known as structure-conditioned protein sequence design, aims to discover protein sequences that can fold into the given structures~\citep{dauparas2022robust,hsu2022learning}. 
Referring to the experimental setup in established studies~\cite{wang2024dplm,hsieh2025elucidating}, the CAMEO 2022 and PDB Date Split datasets are used for evaluation. 
Compared to the one-to-one structure prediction, the inverse folding has a one-to-many nature.
There could be multiple distinct amino acid sequences that can fold into a target structure, in addition to its natural sequence. Therefore, rather than calculating the recovery rate of the natural protein sequence, the evaluation should estimate the self-consistency between the target structure and refolded structure of the designed protein sequence~\citep{liu2025protinvtree}. We calculate the \texttt{scRMSD} and \texttt{scTM} with the assistance of ESMFold~\citep{lin2023evolutionary}. 

\begin{table}[t]\footnotesize
\centering
\captionsetup{font=small,skip=3pt}
\caption{Evaluation of Inverse Folding.
}
{
\setlength{\tabcolsep}{3pt}
\begin{tabular}{lcccc}
\toprule
~ & \multicolumn{2}{c}{CAMEO} & \multicolumn{2}{c}{PDB Date Split} \\
\cmidrule(l){2-3} \cmidrule(l){4-5}
Model & scRMSD $ \downarrow $ & scTM $ \uparrow $ & scRMSD $ \downarrow $ & scTM $ \uparrow $ \\
\midrule
MultiFlow & 4.437 ± 5.024 & 0.876 ± 0.151 & 2.138 ± 2.430 & 0.938 ± 0.074 \\ 
ESM3 & 3.944 ± 4.964 & 0.901 ± 0.141 & 2.262 ± 3.090 & 0.940 ± 0.093 \\ 
\midrule
\rowcolor{gray!10} DPLM-2 (150M) & 5.999 ± 7.469 & 0.848 ± 0.175 & 4.002 ± 4.700 & 0.895 ± 0.126 \\ 
DPLM-2 (650M) & 4.659 ± 4.875 & 0.871 ± 0.154 & 3.114 ± 4.034 & 0.911 ± 0.113 \\ 
DPLM-2.1 & 4.304 ± 4.586 & 0.876 ± 0.141 & 2.271 ± 3.606 & 0.927 ± 0.112 \\ 
\midrule
\rowcolor{gray!10} HD-Prot (155M) & 4.637 ± 4.730 & 0.863 ± 0.156 & 2.903 ± 3.683 & 0.919 ± 0.107 \\ 
HD-Prot (670M) & 4.675 ± 4.930 & 0.866 ± 0.151 & 2.871 ± 3.599 & 0.920 ± 0.103 \\ 
\bottomrule
\end{tabular}
}
\vspace{-0.3cm}
\label{tab6}
\end{table}
The performance of \ourname{} and four baseline methods are summarized in Table~\ref{tab6}, with all sampling procedures run five times with different seeds. 
The evaluation results largely align with those in Table~\ref{tab5}, with MultiFlow standing out for its training on synthetic inverse-folding data.
ESM3 performs best among all methods, excelling at completing the remaining multimodal context when sufficient initial information is provided.
Then, \ourname{} performs highly comparable to the DPLM-2 series at both $\sim$150M and $\sim$650M scales. Such completely zero-shot inverse folding results demonstrate that \ourname{} has estimated the joint-distribution of protein sequence-structure sufficiently well. 
Besides, the sampling strategy of \ourname{} is analyzed in Appendix~\ref{asec:temp_cfg_inv_folding}.

%% file: sections/5-discussion.tex
\section{Conclusion}

Multimodal generative pLMs have recently emerged as a popular solution for jointly modeling protein sequences and structures.
However, the majority of existing methods still suffer from the reliance on quantized discrete structure representations.
To this end, we propose a hybrid diffusion protein language model (\ourname{}), which expands a pre-trained sequence-based pLM to understand and generate continuous protein structure information.
The model bridges the discrete-continuous modality gap in multimodal protein modeling and demonstrates the promising potential of using continuous structure tokens within pLMs.
Extensive quantitative and qualitative experiments show that \ourname{} achieves competitive multimodal protein generation performance compared to state-of-the-art multimodal pLMs, while requiring fewer computational resources for the modality extension development.

\section{Limitations and Ethical Considerations}
There are several limitations warranting future investigation. 
First, our work demonstrates that a multimodal pLM with continuous structure tokens achieves competitive performance despite modest scaling and limited exploration of design choices. 
Second, we currently focus on protein backbone structure and protein monomers, yet our \ourname{} framework can naturally extend to the modeling of all-atom structures and protein complexes. Incorporating an all-atom structure tokenizer and training data of protein multimers represents a promising direction for future work.
Third, our work is still limited to computational benchmarking. Validation through real-world scientific discovery represents a distinct research phase beyond the scope of this study.

In this study, all data used were obtained from publicly available sources, and no ethical approval was required.

\section*{GenAI Disclosure}
The authors used a GenAI solely as a tool to assist with polishing and refining the writing in this paper.
The model was used exclusively to improve grammatical fluency, sentence structure, and the overall clarity of the manuscript.
All ideation, theoretical development, empirical research, and technical conclusions remain entirely the work of the authors.
The authors take full responsibility for all content generated by the GenAI and presented in this work.

%% file: sections/6-appendix.tex
\newpage
\paragraph{Table of Appendix:}
\begin{itemize}
    \item \ref{asec:tokenizer}. Analysis of Continuous Structure Tokens
    \item \ref{app:further explanation}. Further Explanation of HD-Prot
    \begin{itemize}
        \item \ref{asec:schedulers}. Multimodal Protein Modeling
        \item \ref{asec:cogen}. Multimodal Protein Generation Procedure
        \item \ref{asec:cfg}. Classifier-Free Guidance for Continuous Structure Tokens
    \end{itemize}
    \item \ref{app:imp_details}. Implementation Details
    \begin{itemize}
        \item \ref{asec:training_dataset}. Training Dataset
        \item \ref{asec:training}. Training Process of \ourname{}
        \item \ref{asec:baselines}. Implementation of Baseline Models
        \item \ref{asec:metrics}. Metrics Calculations
    \end{itemize}
    \item \ref{app:further analysis}. Further Analysis of Experimental Results
    \begin{itemize}
        \item \ref{asec:uncon_cogen}. Evaluation of Unconditional Sequence-Structure Co-Generation
        \item \ref{asec:cost}. Explanation of Development Cost
        \item \ref{asec:infer_efficiency}. Inference Efficiency Analysis
        \item \ref{asec:scaffolding_problems}. Motif-Scaffolding Results of Each Problem
        \item \ref{asec:mode_analysis}. Co-Generation Cases \& Failure Mode Analysis
        \item \ref{asec:temp_cfg}. Analysis of Sampling Hyperparameters
    \end{itemize}
\end{itemize}

\section{Analysis of Continuous Structure Tokens}
\label{asec:tokenizer}

\begin{figure}[h]
    \centering
    \includegraphics[width=0.9\linewidth]{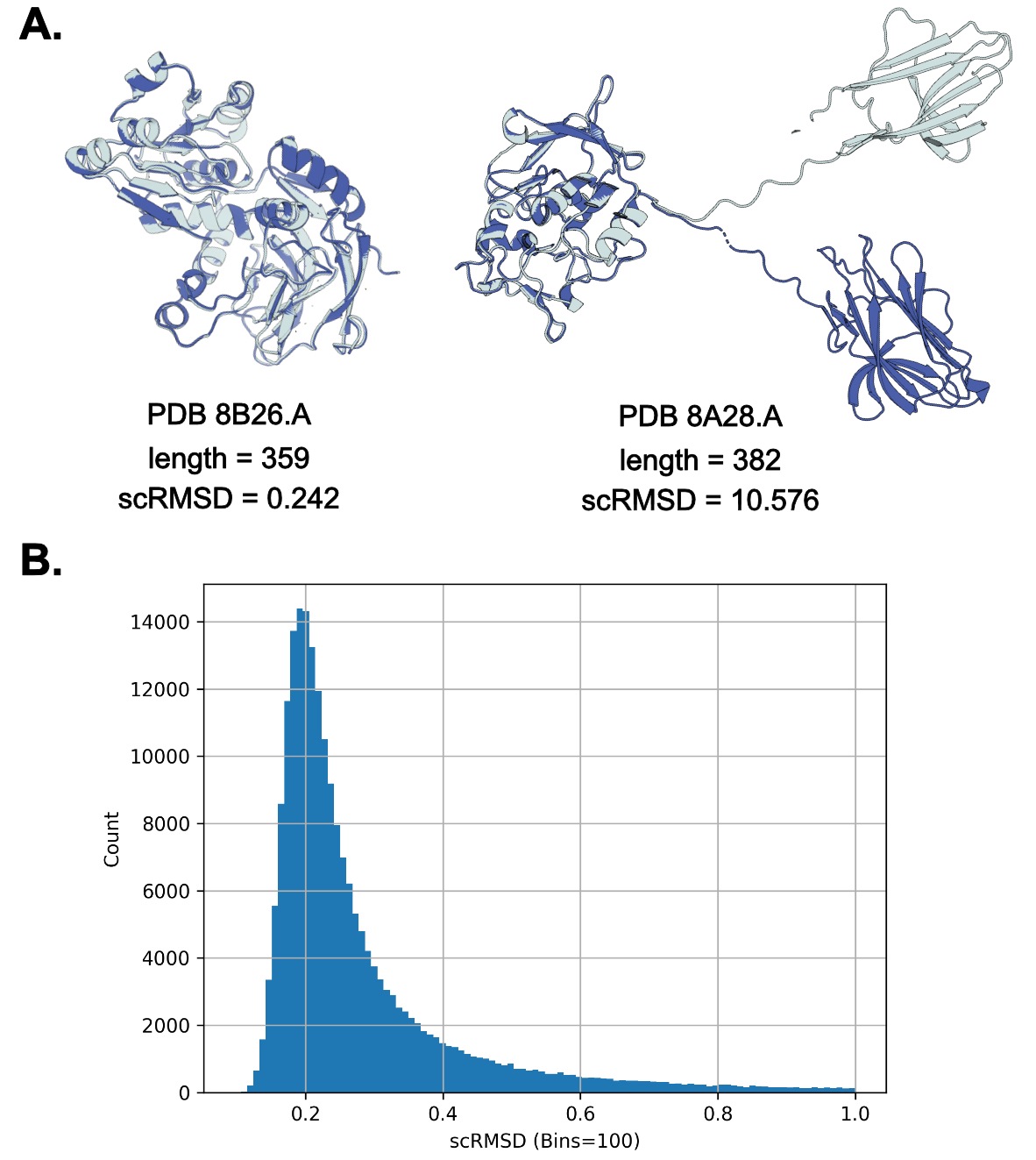}
    \captionsetup{font=small,skip=5pt}
    \caption{Analysis of Continuous Structure Tokens.
    (\textbf{A}) Visualization of protein structure reconstructions. 
    (\textbf{B}) Statistics of the fidelity of continuous structure tokens.}
    \label{afig:recon}
    \vspace{-0.32cm}
\end{figure}

The salad autoencoder~\citep{jendrusch2025efficient} demonstrates excellent performance on the CAMEO 2022 test set, achieving high-fidelity reconstruction with scRMSD $ < $ 1.0 Å for 173 out of 183 test structures.
Figure~\ref{afig:recon}.A presents a random case and a selected bad case, demonstrating the capability and characteristics of the tokenizer. 
It is observed that while the tokenizer achieves consistently accurate \emph{local} reconstructions, it may misorient structural elements in disordered regions, thereby compromising the global performance.

As described in Section~\ref{asec:training_dataset}, our training set contains approximately 210K proteins after various filtering steps.
We pre-cache all those proteins into arrays of continuous structure tokens, and Figure~\ref{afig:recon}.B presents the statistics of the structure reconstruction results based on these token arrays, reflecting their representational fidelity. 
The median scRMSD of 0.229 Å indicates excellent reconstruction quality, demonstrating that continuous structure tokens provide an extensively effective and nearly loss-free representation of protein structures.

While we keep the tokenizer frozen for computational efficiency, we have to adapt to its inherent numerical characteristics. Throughout our training dataset, the numerical mean value of continuous structure tokens is -0.432, and the variance is 28.562. In order to ensure the effective operation of the subsequent continuous diffusion learning based on Gaussian noise $ \epsilon \sim \mathcal{N}(0,\bm{I}) $, all continuous structure tokens undergo a very simple numerical scaling~\citep{li2024autoregressive}. 
Using the statistical mean of the standard deviation as the scaling factor, the numerically divided tokens serve as ground truth for model learning, while the tokens generated by the model are scaled up accordingly for decoding by the tokenizer.

\section{Further Explanation of \ourname{}}
\label{app:further explanation}

\subsection{Multimodal Protein Modeling}
\label{asec:schedulers}

Previous studies~\citep{campbell2024generative,wang2024dplm,li2024speak} have demonstrated that utilizing decoupled sequence and structure diffusion schedulers enables multimodal protein models to achieve comprehensive and fundamental protein modeling.
Table~\ref{tab:schedulers} summarizes the scheduler configurations and their corresponding protein modeling tasks.
We denote the sequence scheduler as $t_s$ and the structure scheduler as $t_z$, where $t_s = 0$ or $t_z = 0$ represents the state of original clean data, and $t_s = T$ or $t_z = T$ corresponds to fully noised data. 

\begin{table}[h]\footnotesize
\centering
\vspace{-0.15cm}
\captionsetup{font=small,skip=3pt}
\caption{Scheduler Settings and Protein Modeling Tasks}
{
\setlength{\tabcolsep}{3.8pt}
\begin{tabular}{cccc}
\toprule
& Sequence Scheduler & Structure Scheduler & Protein Modeling Task \\
\midrule
1 & $ t_s\in \{ 0, 1, \dots, T \} $ & $t_z = T$ & Sequence Generation \\
2 & $t_s = T$ & $ t_z\in \{ 0, 1, \dots, T \} $ & Structure Generation \\
3 & $t_s = 0$ & $ t_z\in \{ 0, 1, \dots, T \} $ & Structure Prediction \\
4 & $ t_s\in \{ 0, 1, \dots, T \} $ & $t_z = 0$ & Inverse Folding \\
5 & \multicolumn{2}{c}{$ t_s = t_z \in \{ 0, 1, \dots, T \} $} & Co-Generation \\
\bottomrule
\end{tabular}
}
\label{tab:schedulers}
\vspace{-0.15cm}
\end{table}

On the one hand, keeping one modality fully masked ensures independent generative modeling of the other modality. By configuring the schedulers as specified in rows 1 and 2 of Table~\ref{tab:schedulers}, the model learns to perform protein sequence generation and protein structure generation, respectively.
On the other hand, maintaining one modality fully visible drives the conditional generation of the other modality. 
The configuration in row 3 enables the model to learn sequence-conditioned structure generation, i.e., protein structure prediction. Similarly, the setting in row 4 facilitates structure-conditioned sequence generation, commonly known as inverse folding. 
Ultimately, by setting $ t_s = t_z \in \{ 0, 1, \dots, T \} $, the model learns sequence-structure dependencies across all possible masking ratios, thereby enhancing protein sequence-structure co-generation.

In the implementation of \ourname{}, we train our model with a combination of three scheduler settings, namely the sequence generation, structure generation, and sequence-structure co-generation. 
In each training batch, one-fifth of samples are treated with $ t_s\in \{ 0, 1, \dots, T \} $ and $t_z = T$ to help the pre-trained sequence-based pLM retain its sequence knowledge. 
Another one-fifth of samples are processed with $t_s = T$ and $ t_z\in \{ 0, 1, \dots, T \} $ to facilitate learning of the newly introduced protein structure modality.  
The remaining 60\% of protein samples are processed with $ t_s = t_z \in \{ 0, 1, \dots, T \} $, enabling the model to learn the joint probability distribution of sequence and structure under positionally interlaced cross-modal conditioning. 
Interestingly, during the explicit training of protein sequence-structure co-generation, \ourname{} also implicitly learn to perform protein structure prediction and inverse folding. We believe that the model, having learned the underlying principles of sequence-structure mapping at the token level, can apply them to complete tracks.

\subsection{Multimodal Protein Generation Procedure}
\label{asec:cogen}

Firstly, we present the most basic procedure of \textit{\textbf{unconditional protein sequence-structure co-generation}} in Algorithm~\ref{alg:cogen}. 
It primarily undergoes the reverse process of diffusion language modeling, i.e., iterative mask token prediction in parallel for both sequence and structure tracks.
Concretely, in each iteration step, discrete sequence tokens are sampled from a categorical distribution, while continuous structure tokens are generated through the reverse process of Denoising Diffusion Probabilistic Model (DDPM)~\citep{ho2020denoising,li2024autoregressive}.
In the end, all generated sequence and structure tokens are translated back to the residue types and 3D coordinates by the tokenizers.

In practice, many extensions can be made to this basic cogeneration process. 
Specific to the sequence track, we adopt some designs native to the foundation model DPLM~\citep{wang2024diffusion}. 
During the sampling of sequence tokens (Algorithm~\ref{alg:cogen} row 11), a resampling scheme is selectively included to prevent the generation of a large proportion of repetitive amino acids.
Meanwhile, instead of the naive random unmasking (row 10), the top-$k$ unmasking strategy selects $k$ tokens with the highest sampling probability score for unmasking. 
Additionally, during the sampling of structure tokens, classifier-free guidance (CFG)~\citep{ho2022classifier} is introduced to enhance sequence-structure self-consistency alongside noise estimation (row 23), with detailed operations described in \ref{asec:cfg}.
Among those sampling hyperparameters, we by default set the diffusion LM steps $ T = L $, the sequence sampling temperature $ \tau_s = 1.0 $, structure DDPM steps $ T' = 100 $, and the DDPM schedule $ \beta_{t'} $ as a linear schedule. 
The trade-off between the self-consistency and diversity of generation results is largely controlled by the structure sampling temperature $ \tau_z $ and the CFG scale. For \ourname{} (155M), our default setting is $ \tau_z =0.35 $ and CFG scale $ =2.0 $. For \ourname{} (670M), the empirically best setting is $ \tau_z =0.55 $ and CFG scale $ =2.0 $.

\begin{algorithm*}[h!]
\captionsetup{}
	\caption{Unconditional Protein Sequence-Structure Co-Generation.}
	\label{alg:cogen}
    \BlankLine
    \KwIn{\begin{tabular}{@{}ll}
        $\quad$Network: & Trained network $\theta = (\theta_b, \theta_s, \theta_z)$ (backbone, categorical head, denoising head) \\
        $\quad$Hyperparams: & Desired protein length $L$; Diffusion language modeling steps $T$ \\
        $\quad$Sequence track: & Sampling temperature $\tau_s$ \\
        $\quad$Structure track: & Sampling temperature $\tau_z$; DDPM steps $T'$; DDPM schedule ${\beta_{t'}}$
    \end{tabular}}
    \KwOut{$\,$Generated protein $(\bm{s}^{(0)}, \bm{z}^{(0)})$}
	\BlankLine

    \BlankLine
    \textbf{Initializations:} \\
    \BlankLine
    \For{$ i = 1, 2, \dots, L $}{
        $\bm{s}_i^{(T)} \gets \bm{m}_s$, $\bm{z}_i^{(T)} \gets \bm{m}_z$; 
        \Comment{Initialize all tokens with masks}
    }
    $k \gets \left\lfloor L / T \right\rfloor$;
    \Comment{Number of tokens to update in each of the following steps} \\
    \BlankLine
    \BlankLine
    \textbf{Reverse Process of Diffusion Language Modeling:} \\
    \BlankLine
	\For{$ t = T, \dots, 1 $}{
        \BlankLine
        $\bm{c}^{(t)} \gets f_{\theta_b}(\bm{s}^{(t)}, \bm{z}^{(t)})$; 
        \Comment{Inference through the main body of pLM} \\
        \BlankLine
        \BlankLine
        \textbf{Sequence Track Update:} \\
        \BlankLine
        $\hat{\bm{s}}^{(0)} \sim \text{Softmax}(f_{\theta_s}(\bm{c}^{(t)}) / \tau_s)$; 
        \Comment{Sample sequence tokens from a categorical distribution} \\
        $\mathcal{I}_{\text{s}}^{(t)} \gets \text{RandomSelect}\left(k, \{i \mid \bm{s}^{(t)}_i = \bm{m}_s\}\right)$; 
        \Comment{Randomly select $k$ \textit{masked} tokens to update} \\
        \For{$i = 1$ \KwTo $L$}{
            \uIf{$i \in \mathcal{I}_{\text{s}}^{(t)}$}{
                $\bm{s}^{(t-1)}_i \gets \hat{\bm{s}}^{(0)}_i$;
                \Comment{Update with newly sampled sequence token}
            }
            \Else{
                $\bm{s}^{(t-1)}_i \gets \bm{m}_s$;
                \Comment{Keep the other sequence tokens masked}
            }
        }
        
        \BlankLine
        \BlankLine
        \textbf{Structure Track Update:} \\
        \BlankLine
        $ \hat{\bm{z}}^{(T')} \sim \mathcal{N}(0,\bm{I})$; 
        \Comment{Sample continuous structure tokens starting from Gaussian noise} \\
        \For{$t' = T', T'-1, \dots, 1$}{
            $\alpha_{t'} := 1 - \beta_{t'}$, \quad $\bar{\alpha}_{t'} := \Pi_{n=1}^{t'} \alpha_n$, \quad $\sigma^2 = \beta_{t'}$, \quad $\delta \sim \mathcal{N}(0,\bm{I})$; 
            \Comment{Setup the DDPM scheduler and additional random noise} \\
            $ \hat{\epsilon} \gets \epsilon_{\theta_z}(\hat{\bm{z}}^{(t')}, t', \bm{c}^{(t)}) $; \Comment{Noise prediction} \\
            $ \hat{\bm{z}}^{(t'-1)} \gets \frac{1}{\sqrt{\alpha_{t'}}} \left( \hat{\bm{z}}^{(t')} - \frac{1-\alpha_{t'}}{\sqrt{1-\bar{\alpha}_{t'}}} \hat{\epsilon} \right) + (\sigma_{t'} \delta)\tau_z $; 
            \Comment{DDPM denoising step} \\
        }
        $\mathcal{I}_{\text{z}^{(t)}} \gets \text{RandomSelect}\left(k, \{i \mid \bm{z}^{(t)}_i = m_z\}\right)$; \Comment{Randomly select $k$ \textit{masked} tokens to update} \\
        \For{$i = 1$ \KwTo $L$}{
            \uIf{$i \in \mathcal{I}_{\text{z}}^{(t)}$}{
                $\bm{z}^{(t-1)}_i \gets \hat{\bm{z}}^{(0)}_i$;
                \Comment{Update with newly sampled structure tokens}
            }
            \Else{
                $\bm{z}^{(t-1)}_i \gets \bm{m}_z$;
                \Comment{Keep the other tokens structure masked}
            }
        }
        \BlankLine
    }

    \BlankLine
	\textbf{return} $ ( \bm{s}^{(0)}, \bm{z}^{(0)} ) $; \Comment{Return the generated protein}
    \BlankLine
\end{algorithm*}

In addition to unconditional sequence-structure co-generation, \ourname{} is also capable of conditional co-generation, i.e., \textit{\textbf{motif-scaffolding}}. We only need to modify the input initialization. 
Different from rows 1-5 of Algorithm~\ref{alg:cogen}, we don't initialize all tokens with masks. 
Given a motif with a length of $ l $, its sequence is directly mapped to the sequence tokens, and its structure is first processed into continuous structure tokens via the tokenizer.
According to the specific motif position and scaffold length $ L $~\citep{yim2024improved}, the input consists of the sequence and structure tokens of the motif at their specific positions, while other positions are masked. \ourname{} gradually generates all mask tokens over $ T = L - l $ steps, with the initialized motif tokens maintain unchanged. 
Following the final step, all sequence and structure tokens are translated back to the residue types and 3D coordinates by the tokenizers.
For both \ourname{} (155M) and \ourname{} (670M), we by default set the sequence sampling temperature $ \tau_s = 1.0 $ and the structure sampling temperature $ \tau_z = 0.1 $, without using the classifier-free guidance.

To accomplish the \textit{\textbf{protein structure prediction}}, the sequence track of \ourname{} is initialized according to the given protein sequence, and the structure track is completely filled with mask tokens. 
No matter what the length of the given protein sequence is, \ourname{} predicts all structure tokens in one step, i.e., setting $ T = 1 $. By default, the structure sampling temperature $ \tau_z = 0.0 $, without employing the classifier-free guidance.
Subsequently, the generated continuous structure tokens are transformed into 3D coordinates via the structure tokenizer.
Similarly, for \textit{\textbf{inverse folding}}, a given protein structure with length $ L $ is firstly processed into the continuous structure tokens by our protein structure tokenizer. Then, the structure track of \ourname{} is initialized by those structure tokens, and the sequence track is set as fully masked. 
By default, \ourname{} gradually predicts all sequence tokens over $ T = L $ steps with sequence sampling temperature $ \tau_s = 0.1 $. The finally obtained sequence tokens are mapped to the amino acid sequence.

Consistent with existing multimodal pLMs~\citep{hayes2025simulating,wang2024dplm}, \ourname{} requires specific sampling strategies for different multimodal protein generation tasks. 
In Section~\ref{asec:temp_cfg}, we present further ablation studies on the selection of sampling hyperparameters. 
It is observed that the optimal sampling strategy depends on the strength of the given condition. 
Stronger conditions constrain the model to a narrower solution space, allowing it to achieve better performance with lower temperature, fewer diffusion steps, and without guidance.

\subsection{Classifier-Free Guidance for Continuous Structure Tokens}
\label{asec:cfg}

Classifier-free guidance~\citep{ho2022classifier} has been extensively utilized in diffusion generative models.
For example, in vision models and vision language models, CFG is commonly used to generate high-quality images that align better with the condition labels or prompts~\citep{li2024autoregressive,wu2025janus}. 
The core idea of CFG is to extrapolate the model's output by combining a conditional prediction and an unconditional prediction, steering the generation towards the condition by increasing the scale of the difference between them. It concretely adjusts the noise estimation of diffusion models through:
\begin{equation}
    \hat{\epsilon} \gets \left(1-\omega\right) \cdot \underbrace{\epsilon_{\theta}\left(\bm{x}\mid\emptyset\right)}_{\text{unconditional}} + \ \omega \cdot  \underbrace{\epsilon_{\theta}\left(\bm{x}\mid\bm{c}\right)}_{\text{conditional}},
\end{equation}
where $ \bm{x} $ denotes the model's input general input content, $ \bm{c} $ denotes the generation condition, and $ \omega $ is the guidance scale.

Our \ourname{} framework fuses the sequence and structure information from the very beginning. Any change in the input sequence/structure is bound to have an impact on the output structure/sequence.
Therefore, the unconditional sequence-structure co-generation process can be treated as $T$-step combination of cross-modal conditional generation of tokens. 
Specifically, we consider the whole sequence track as the condition for the sampling of continuous structure tokens, where fully masking the sequence track is a kind of ``unconditional" case.

The DDPM generation process for continuous structure tokens naturally supports classifier-free guidance. Introducing the conditional and unconditional cases that we just explained, CFG changes the noise prediction described in row 23 of Algorithm~\ref{alg:cogen}, formally expressed as:
\begin{equation}
    \hat{\epsilon} \gets \left(1-\omega\right) \cdot \underbrace{\epsilon_{\theta_z}(\hat{\bm{z}}^{(t')}, t', \bm{c}_{\emptyset}^{(t)})}_{\text{unconditional}} \ + \  \omega \cdot  \underbrace{\epsilon_{\theta_z}(\hat{\bm{z}}^{(t')}, t', \bm{c}^{(t)})}_{\text{conditional}},
\end{equation}
where $ \bm{c}_{\emptyset}^{(t)} \gets f_{\theta_b}(\emptyset, \bm{z}^{(t)}) $ involves an additional inference through the pLM with sequence tokens fully masked, and $ \omega $ is the CFG scale. 

\section{Implementation Details}
\label{app:imp_details}

\subsection{Training Dataset}
\label{asec:training_dataset}

A well-constructed training dataset plays an essential role in the successful training of protein generative models. 
Accordingly, various ``AI for Protein" projects have designed specific schemes to cluster and filter experimental and synthetic data from PDB~\citep{wwpdb2019protein} and AlphaFoldDB~\citep{varadi2022alphafold}.
Our dataset is built on the DPLM-2 practice~\citep{wang2024dplm}, which comprises approximately 20K PDB proteins and 200K APDB-SwissProt proteins. 
The former are representative clustering centers of PDB monomer proteins determined before Aug 2021, and the latter are high-quality protein structure predictions with $\text{pLDDT} > 85$. 
Rather than directly using their data processing results, we independently obtain all protein structures based on the protein name list and perform additional filtering for structure reconstruction quality. 
We hypothesize that if a structure can not be excellently encoded and reconstructed by our protein structure tokenizer~\citep{jendrusch2025efficient}, it should be misleading to learn the probability distribution of the corresponding continuous structure tokens. 
By requiring a structure reconstruction quality of $\text{scRMSD} < 1.0$ and $\text{scTM} > 0.9$, our dataset ultimately comprises 210,001 samples, consisting of 19,807 PDB proteins and 190,194 AFDB-SwissProt proteins. 

\begin{figure}[h]
  \centering
  \includegraphics[width=0.8\linewidth]{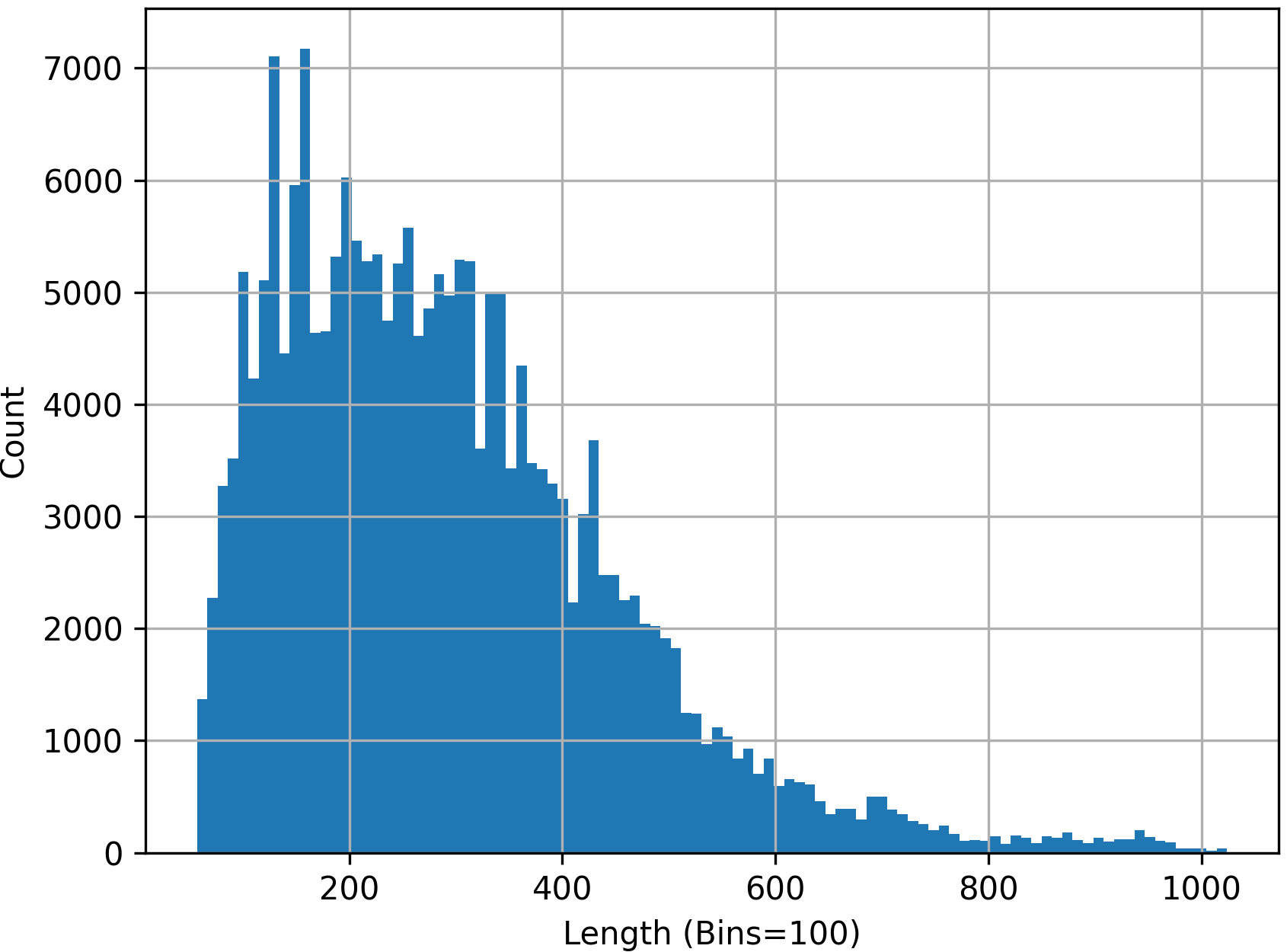}
  \captionsetup{font=small,skip=6pt}
  \caption{Length statistics of the training proteins}
  \label{afig:seq_len}
  \vspace{-0.3cm}
\end{figure}

As shown in Figure~\ref{afig:seq_len}, the proteins in our training dataset have lengths ranging from 57 to 1024 residues. 
During the model's training, proteins longer than 512 residues are randomly cropped to a length between 384 and 512. 
Furthermore, a random cropping strategy~\citep{wang2024dplm} is also introduced to enhance data diversity. 
Any protein with more than 60 residues has a 50\% chance of being cropped to a random length between 60 and its full length.

\subsection{Training Process of \ourname{}}
\label{asec:training}

We employ the protein structure tokenizer with its pretrained parameters frozen. All training data are preprocessed into paired discrete sequence tokens and continuous structure tokens, then cached for efficient access. 
DPLM~\citep{wang2024diffusion}, a pretrained sequence-based protein language model, is adopted as the foundation model. 
\ourname{} (155M) initializes its pLM backbone from DPLM (150M), and similarly, \ourname{} (670M) is initialized from DPLM (650M). 
Overall, the trainable parameters include the pLM backbone (fine-tuned) and the remaining modules (trained from scratch). 
According to our existing empirical observations, we recommend different training strategies for the two model scales: full-model fine-tuning for the 150M backbone, and a LoRA~\citep{hu2022lora} configuration that yields $\sim$91M trainable parameters for the 650M backbone.

For hyperparameters, we adopt the reweighting scheme from DPLM~\cite{wang2024diffusion} for the sequence track, setting $\lambda^{(t_s)} = 1 - (t_s - 1) / T$. For the structure track, we maintain a constant weight of $\lambda^{(t_z)} = 1$ following MAR~\cite{li2024autoregressive}, and the DDPM diffusion schedule $\beta_{t'}$ is simply a linear schedule. The $\gamma$ used to combine the sequence/structure modeling losses is set as $0.2$ empirically, aiming to balance the magnitudes of the two loss values. 
For optimization, we use AdamW optimizer~\citep{loshchilov2017decoupled} with $ \beta_1 = 0.9 $, $ \beta_2 = 0.95 $ and the weight decay $ =0.01 $. 
Mixed-precision technique is also introduced to reduce memory consumption.
The training of \ourname{} runs for 120 epochs: warmup from 1e-5 to 1e-4 over the first 5 epochs, and linear decay to 1e-5 over the other 115 epochs. 
\ourname{} (155M) takes \textbf{1} NVIDIA H20-96G GPU for approximately 7 days, and \ourname{} (670M) takes \textbf{2} NVIDIA H20-96G GPUs for about 10 days.

\subsection{Implementation of Baseline models}
\label{asec:baselines}

We run MultiFlow~\citep{campbell2024generative}, La-Proteina~\citep{geffner2025laproteina} and PLAID~\citep{lu2025all} using their official checkpoints and pipelines.\footnote{https://github.com/jasonkyuyim/multiflow}\footnote{https://github.com/NVIDIA-Digital-Bio/la-proteina}\footnote{https://github.com/amyxlu/plaid}
For La-Proteina, we evaluate both variants (with/without triangular updates) with the default noise scales of 0.1 for the alpha carbon atoms and 0.1 for the latent variables. 
For PLAID, we evaluate both the 100M- and 2B-parameter models. The lengths of proteins sampled with PLAID are $ \{ $104, 200, 304, 400, 504$ \} $, as it supports only lengths divisible by 8.

DPLM-2 series~\citep{wang2024dplm,hsieh2025elucidating} are also implemented by using their official checkpoints following the latest official instructions.\footnote{https://github.com/bytedance/dplm}
For unconditional sequence-structure co-generation and motif-scaffolding, DPLM-2 employs default sampling strategies of ``annealing@2.0:0.1" and ``annealing@2.0:1.0", respectively, both over 500 steps. For protein structure prediction and inverse folding, DPLM-2 instead performs argmax sampling for 100 steps.
DPLM-2.1 by default adopts the ``annealing@1.1:0.1" strategy for unconditional co-generation over 500 steps, and similarly uses argmax sampling for both protein structure prediction and inverse folding.

Notably, the ESM3~\citep{hayes2025simulating} official\footnote{https://github.com/evolutionaryscale/esm}
provides the pre-trained checkpoint but has not specified how to perform unconditional sequence-structure co-generation. We adopt the suggestions of~\citet{yim2025hierarchical} to perform a chain-of-thought inference to generate protein backbone structures first, including the sampling of secondary structure tokens with a temperature of 0.7, followed by the sampling of structure tokens with a temperature of 0.7. Subsequently, we sample the corresponding protein sequences at a temperature of 0.7. The three consecutive sets of sampling are all completed in $L$ steps ($L$ is the desired protein length).
We attempted to implement ESM3 using the sequence-structure order instead of the secondary structure-structure-sequence order, or using other temperature settings, but did not achieve better results. 
Moreover, as described in the original appendix, ESM3 employs single-pass argmax decoding for protein structure prediction, and iterative decoding over $L/2$ steps using a constant temperature of 0.7 for inverse folding.
As exemplified in the official tutorial\footnote{https://github.com/evolutionaryscale/esm/blob/main/cookbook/\\tutorials/4\_forge\_generate.ipynb},
for motif-scaffolding, ESM3 iteratively sample a protein sequence based on the motif prompt over $L/2$ steps with the temperature annealing from 0.5, subsequently predict the structure of the generated sequence by iteratively sampling structure tokens over $L/8$ steps with the temperature annealing from 0.7.

\subsection{Metrics Calculations}
\label{asec:metrics}

Throughout all experiments, the \texttt{RMSD} and \texttt{TM-score} are calculated using standard functions in OpenFold~\citep{ahdritz2024openfold} and TM-Tools~\citep{zhang2005tm}.
The \texttt{\#Clusters@50} and \texttt{\#Cluster@95} are obtained by clustering the generated structures pooled by length via Foldseek~\citep{van2024fast}, using the following command:
{\small
\begin{align*}
    &\texttt{foldseek easy-cluster } \langle \texttt{input\_path} \rangle \ \langle \texttt{output\_path} \rangle \ \langle \texttt{tmp\_path} \rangle \\[-1.8pt]
    &\texttt{--alignment-type 1 --cov-mode 0 --min-seq-id 0} \\[-1.8pt]
    &\texttt{--tmscore-threshold 0.5 \# or --tmscore-threshold 0.95} \text{.} 
\end{align*}

}

We quantify novelty by searching each generated protein against a reference database using Foldseek with the command below. The highest TM-score from the alignment against the PDB proteins is recorded as the \texttt{pdb-TM}, and that against AlphaFoldDB-SwissProt proteins as the \texttt{sp-TM}. Lower values indicate greater structural dissimilarity to existing databases.
{\small
\begin{align*}
    &\texttt{foldseek easy-search } \langle \texttt{input\_path} \rangle \ \langle \texttt{database\_path} \rangle \\[-1.8pt]
    &\langle \texttt{output\_path} \rangle \ \langle \texttt{tmp\_path} \rangle \texttt{ --exhaustive-search} \texttt{ --alignment-type 1} \\[-1.8pt] 
    &\texttt{--tmscore-threshold 0.0 --format-output query,target,alntmscore} \text{.}
\end{align*}
}

However, \texttt{pdb-TM} and \texttt{sp-TM} should be interpreted with care: they do not measure novelty in an absolute sense, but rather quantify distance from the chosen reference distributions. 
In unconditional protein generation, model outputs inherently reflect the statistics of the training data. 
Therefore, models trained on datasets more closely aligned with PDB/AlphaFoldDB-SwissProt may obtain worse novelty scores while producing more naturalistic proteins. 
Meanwhile, models trained with more synthetic data generated by ProteinMPNN+ESMFold may appear more novel, simply because their generated structures deviate further from PDB/AlphaFoldDB-SwissProt. 
That is, novelty metrics inherently favor synthetic training data,
which may not always align with the goal of learning fundamental protein sequence-structure principles. 

\section{Further Analysis of Experimental Results}
\label{app:further analysis}

\subsection{Evaluation of Unconditional Sequence-Structure Co-Generation}
\label{asec:uncon_cogen}

\begin{figure*}[t]
    \centering
    \includegraphics[width=0.8\linewidth]{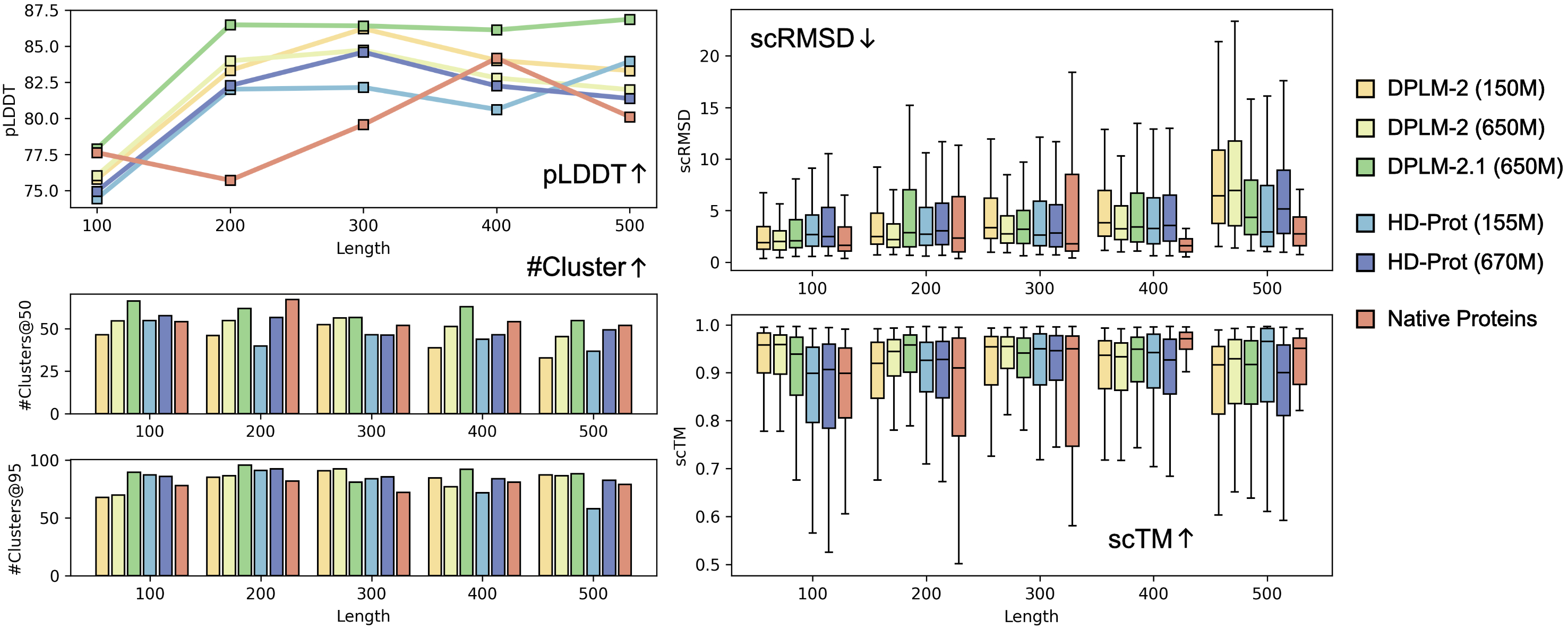}
    \captionsetup{font=small,skip=6pt}
    \caption{Evaluation on Unconditional Sequence-Structure Co-Generation}
    \label{afig:uncon_cogen}
\end{figure*}

Figure~\ref{afig:uncon_cogen} shows the detailed performance of the DPLM-2 series and \ourname{} grouped by the protein length, alongside the characteristics of native proteins for reference. Notably, the foldability, self-consistency, and diversity of native proteins remain largely unaffected by protein length. In contrast, both the DPLM-2 series and \ourname{} produce less self-consistent and more repetitive proteins as the specified length increases. We believe the issue lies with the data. All natural proteins, irrespective of length, are governed by fundamental physical and evolutionary principles that underlie their stable existence. However, the principles have not been explicitly elucidated, and current AI models merely fit them implicitly in a data-driven manner. The limited presence of longer proteins in the training data (Figure~\ref{afig:seq_len}) consequently leads to a drop in generation performance.

\subsection{Explanation of Development Cost}
\label{asec:cost}

Firstly, we briefly summarize the development processes of the baseline methods:
\begin{itemize}[leftmargin=*]
\item MultiFlow~\citep{campbell2024generative} is trained from scratch with 4 NVIDIA A6000 GPUs for 3 days. 
\item La-Proteina~\citep{geffner2025laproteina} is trained from scratch. Its VAE training uses 16, 32, and 32 NVIDIA A100 (80G) GPUs across three stages, followed by flow-matching training with 48, 96, and 128 NVIDIA A100 (80G) GPUs across another three stages.
\item PLAID~\citep{lu2025all} is built on the basis of ESMFold. With ESMFold frozen, PLAID is trained from scratch to learn the shared sequence-structure latent within, requiring 8 NVIDIA A100 GPUs.
\item ESM3~\citep{hayes2025simulating} is developed from scratch, with the protein structure tokenizer and the main model trained separately. Although the paper does not directly report the detailed compute budget, its model scale ($\geq$1.4B parameters) and training data volume ($\geq$1.08B proteins) indicate a substantial training-resource requirement.
\item DPLM-2~\citep{wang2024dplm} first develops a protein structure tokenizer comprising a GVP-based encoder and an IPA-based decoder, with the encoder initialized from GVP-Transformer~\citep{hsu2022learning}. The tokenizer is trained on 8 NVIDIA A100 GPUs for 2 days. The main model is then initialized from DPLM~\citep{wang2024diffusion} and fine-tuned for modality extension with LoRA on 8 or 16 NVIDIA A100 GPUs for 3 days, depending on the model scale.
\item DPLM-2.1~\citep{hsieh2025elucidating} reuses the protein structure tokenizer of DPLM-2 and also initializes the main model from DPLM for modality extension. The specific compute requirement is not reported.
\end{itemize}

Among these baselines, \ourname{} and DPLM-2 are the closest counterparts in model architecture and development procedure, since both extend a pretrained sequence-based pLM, i.e., DPLM, to multimodal protein modeling. 
Table~\ref{atab:comp_resources} compares the estimated development cost of this modality-extension stage.
\begin{table}[h!]\footnotesize
    \centering
    \captionsetup{font=small,skip=3pt}
    \caption{Modality Extension Cost Comparison}
    \begin{tabular}{ccc}
    \toprule
        \multirow{2}{*}{\makecell[c]{Model}} & \multirow{2}{*}{\makecell[c]{Device Requirements\\ (GPU Type $ \times $ Count $ \times $ Day)}} & \multirow{2}{*}{\makecell[c]{Estimated\\ Cost (CNY)}} \\
        \\
        \midrule
        DPLM-2 (150M) & NVIDIA A100 $\times$ 8 $ \times $ 3 & 15,792 \\
        DPLM-2 (650M) & NVIDIA A100 $\times$ 16 $ \times $ 3 & 31,584 \\
        \midrule
        HD-Prot (155M) & NVIDIA H20 $\times$ 1 $ \times $ 7 & 1,036 \\
        HD-Prot (670M) &NVIDIA H20 $\times$ 2 $ \times $ 10 & 2,960 \\
        \bottomrule
    \end{tabular}
    \label{atab:comp_resources}
\end{table}

Our experiments use the NVIDIA H20 GPU, which is only available in certain regions and limited cloud computing platforms due to export controls on high-end AI accelerators. 
The cost estimates are therefore based on the prevailing pricing of available cloud platforms.
On the AutoDL platform, renting one H20 (96G) GPU costs approximately 4,420 CNY per month (148 CNY per day). Training the \ourname{} (670M) model requires 10 days on two H20, leading to an estimated cost of \textbf{\textit{2,960}} CNY. 
For comparison, on the Volcengine platform, renting one A100 (80G) GPU costs about 19718 CNY per month (658 CNY per day). Training the DPLM-2 (650M) model, which required 16 A100 GPUs for 3 days, will cost approximately \textbf{\textit{31,584}} CNY. 
It suggests that our modality extension fine-tuning cost is less than \textbf{\textit{one-tenth}} of that of DPLM-2.

\subsection{Inference Efficiency Analysis}
\label{asec:infer_efficiency}

\begin{table}[h!]\footnotesize
\centering
\captionsetup{font=small,skip=3pt}
\caption{Inference Time Comparisons.}
{
\begin{tabular}{lcccc}
\toprule
~ & \multicolumn{4}{c}{Batch Size} \\
\cmidrule(lr){2-5}
Model & 1 & 10 & 50 & 100 \\
\midrule
ESM3 & 21.4 & - & - & - \\
DPLM-2 (150M) & 15.4 & 33.9 & 137.3 & 267.4 \\
DPLM-2 (650M) & 17.6 & 59.4 & 257.7 & 502.6 \\
DPLM-2.1 (650M) & 16.5 & 52.1 & 219.4 & 425.5 \\
\ourname{} (155M) w/o CFG & 30.1 & 37.2 & 70.0 & 100.9 \\
\ourname{} (670M) w/o CFG & 32.5 & 47.7 & 103.7 & 249.1 \\
\ourname{} (155M) w/ CFG & 34.8 & 58.9 & 153.8 & 160.8 \\
\ourname{} (670M) w/ CFG & 37.2 & 89.4 & 250.8 & 465.5 \\
\bottomrule
\end{tabular}
}
\label{tab:infer_efficiency}
\end{table}

In Table~\ref{tab:infer_efficiency}, we report the average time (in seconds) required to generate proteins of length $L=200$ on a single NVIDIA H20 GPU under both sequential and batched inference settings. From these results, we draw four main observations:

\begin{itemize}[leftmargin=*]
\item Batched sampling makes much better use of GPU parallelism and is substantially more efficient than sequential generation, although not all methods support batched inference (e.g., ESM3).
\item Although HD-Prot is slower in the sequential setting due to limited GPU utilization, it achieves inference efficiency comparable to, or even better than, DPLM-2/2.1 under batched settings.
\item Inference efficiency is influenced by multiple factors. From an architectural perspective, \ourname{} uses a DDPM-based denoising head for continuous structure generation, which is less efficient than a standard categorical prediction head. However, its sequence-structure summation fusion is computationally more efficient than concatenation-based fusion in the pLM backbone, operating on a context length $L$ rather than $2L$. In addition, the sampling strategy, especially the total number of denoising steps and the use of CFG, also has a substantial impact on efficiency.
\item As a direction for future work, replacing DDPM in the structure denoising head with flow matching may further improve the inference efficiency of \ourname{}.
\end{itemize}

\subsection{Motif-Scaffolding Results of Each Problem}
\label{asec:scaffolding_problems}

Table~\ref{tab:scaffolding_problems} details the motif-scaffolding results. 
We have conducted five repetitions using five different random seeds. We summarize the average, minimum, and maximum number of times each problem is solved, and report the average success rate with standard deviation.

\begin{table}[h!]\footnotesize
\centering
\captionsetup{font=small,skip=3pt}
\caption{Motif-Scaffolding Results of Each Problem. }
{
\setlength{\tabcolsep}{2.3pt}
\begin{tabular}{cccccc}
\toprule
& \multirow{2}{*}{ESM3} & \multirow{2}{*}{\makecell[c]{DPLM-2\\ (150M)}} & \multirow{2}{*}{\makecell[c]{DPLM-2\\ (650M)}} & \multirow{2}{*}{\makecell[c]{\ourname{}\\ (155M)}} & \multirow{2}{*}{\makecell[c]{\ourname{}\\ (670M)}} \\
\\
\midrule
1BCF & 87.4 (81, 98) & {6.4 (4, 10)} & {0.8 (0, 2)} & {5.4 (1, 9)} & {9.6 (7, 14)} \\
1PRW & 94.4 (89, 98) & {88.8 (87, 91)} & {80.2 (76, 85)} & {70.4 (62, 79)} & {78.8 (74, 82)} \\
1QJG & 0.8 (0, 3) & 0 & 0 & 0 & 0 \\
1YCR & 60.0 (55, 66) & {29.2 (25, 33)} & {38.2 (34, 46)} & {44.2 (37, 53)} & {45.2 (36, 61)} \\
2KL8 & 0 & {44.2 (39, 54)} & {64.2 (58, 76)} & {50.4 (46, 58)} & {59.0 (55, 63)} \\
3IXT & 51.6 (44, 59) & {36.4 (32, 42)} & {53.6 (44, 74)} & {51.4 (48, 57)} & {38.0 (33, 49)} \\
4JHW & 0.2 (0, 1) & 0 & 0 & 0 & 0 \\
4ZYP & 26.2 (16, 32) & {4.8 (3, 6)} & {11.6 (7, 15)} & {0.4 (0, 1)} & {2.0 (1, 3)} \\
5IUS & 0 & 0 & 0 & 0 & {0.2 (0, 1)} \\
5TPN & 22.0 (17, 28) & {0.4 (0, 1)} & {0.4 (0, 1)} & {15.2 (12, 20)} & {11.8 (6, 15)} \\
5TRV\_long & 31.2 (25, 39) & {2.2 (1, 5)} & {1.6 (0, 3)} & {8.6 (8, 10)} & {8.6 (3, 13)} \\
5TRV\_med & 19.4 (12, 27) & {6.2 (4, 10)} & {6.6 (4, 9)} & {11.4 (8, 15)} & {20.0 (17, 25)} \\
5TRV\_short & 3.2 (2, 5) & {0.8 (0, 2)} & {1.6 (1, 3)} & {10.4 (6, 16)} & {17.4 (12, 23)} \\
5WN9 & 8.2 (5, 11) & {0.2 (0, 1)} & 0 & 0 & {0.2 (0, 1)} \\
5YUI & 0 & 0 & 0 & 0 & 0 \\
6E6R\_long & 17.8 (9, 23) & {70.2 (68, 72)} & {69.8 (65, 75)} & {13.4 (9, 19)} & {24.0 (18, 30)} \\
6E6R\_med & 33.6 (30, 39) & {53.0 (50, 56)} & {65.0 (61, 71)} & {18.2 (16, 21)} & {27.8 (25, 30)} \\
6E6R\_short & 41.0 (33, 51) & {52.8 (50, 54)} & {64.8 (62, 69)} & {35.2 (28, 42)} & {49.4 (37, 57)} \\
6EXZ\_long & 18.8 (14, 23) & {30.6 (23, 37)} & {53.6 (49, 60)} & {6.6 (4, 10)} & {36.0 (25, 39)} \\
6EXZ\_med & 27.6 (23, 32) & {32.8 (30, 35)} & {51.4 (46, 58)} & {8.6 (5, 11)} & {37.6 (28, 46)} \\
6EXZ\_short & 32.4 (28, 36) & {20.0 (10, 24)} & {28.8 (46, 58)} & {12.2 (7, 16)} & {52.0 (42, 61)} \\
7MRX\_128 & 44.8 (41, 49) & 0 & {15.4 (6, 23)} & {1.8 (0, 3)} & {8.4 (5, 17)} \\
7MRX\_60 & 64.8 (57, 71) & {0.6 {(0, 2)}} & {30.4 (25, 38)} & {9.2 (5, 13)} & {31.6 (30, 33)} \\
7MRX\_85 & 61.4 (59, 63) & 0 & {26.0 {22, 32}} & {7.4 (5, 11)} & {21.6 (16, 26)} \\
\midrule
$\#$Solved{/24} & 19.6 (19, 20) & {15.6 (14, 17)} & {17.8 (16, 19)} & {18.2 (18, 19)} & {19.4 (19, \textbf{21})} \\
Success & \textbf{30.1\%} ± 0.3\% & {20.0\% ± 7.0\%} & {27.7\% ± 0.8\%} & {15.8\% ± 0.3\%} & {24.1\% ± 1.1\%} \\
\bottomrule
\end{tabular}
}
\label{tab:scaffolding_problems}
\vspace{-0.3cm}
\end{table}

\subsection{Co-Generation Cases \& Failure Mode Analysis}
\label{asec:mode_analysis}

\begin{figure}[t]
    \centering
    \includegraphics[width=\linewidth]{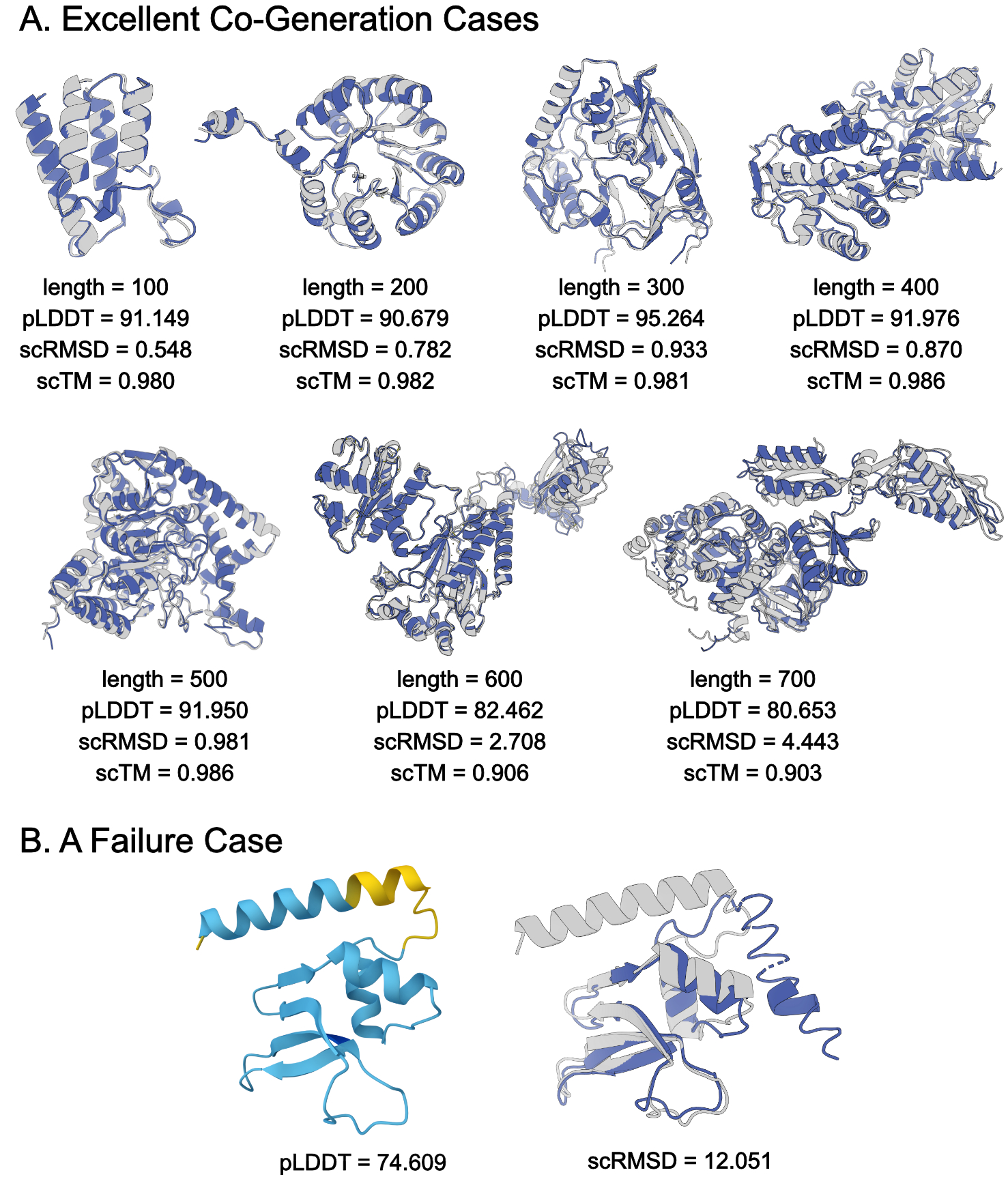}
    \captionsetup{font=small,skip=6pt}
    \caption{Case study for unconditional sequence-structure co-generation. (\textbf{A}) Successful examples. In the structure alignment visualizations, co-generated structures are shown in blue and ESMFold-predicted structures in gray. (\textbf{B}) A failure case. The ESMFold-predicted structure is colored by the pLDDT scores, where the light blue indicates $ 70 < \texttt{pLDDT} < 90 $ and the yellow indicates $ 50 < \texttt{pLDDT} < 70 $. It is also aligned with and compared to the co-generated structure.}
    \label{afig:cases}
    \vspace{-0.3cm}
\end{figure}

Fugure \ref{afig:cases}.A presents some excellent protein sequence-structure co-generation cases produced by \ourname{}. The selected examples, with lengths of 100-500, exhibit a high degree of foldability (\texttt{pLDDT} $> 90$) and self-consistency (\texttt{scRMSD} $< 1.0$, \texttt{scTM} $> 0.9$). Meanwhile, although our model was trained primarily on proteins shorter than 512 residues, we can still find certain good cases for larger proteins with 600/700 residues.

Additionally, we analyze a representative failure case, visualized in Figure \ref{afig:cases}.B. 
In this example, ESMFold reports a relatively high global folding confidence (mean pLDDT $= 74.609$). However, a closer inspection of per-residue pLDDT scores reveals a short coil segment that connects an alpha-helix to the rest of the structure with markedly lower confidence ($50 < \text{pLDDT} < 70$), indicating uncertainty in the helix’s precise orientation. 
We posit two plausible explanations: either the sequence generated by \ourname{} is suboptimal, or the region corresponds to a genuine disordered segment.
Visualization of the protein structure alignment further shows that the alpha-helix is oriented in markedly different directions in the co-generated and ESMFold-predicted structures, leading to a poor RMSD score.
Moreover, the alpha-helix in the co-generated structure exhibits unphysical distortions, suggesting low structural rationality. 
We posit that the corresponding continuous structure tokens remain noisy.
Overall, we identify two common error patterns: 1) the structure orientation is misjudged when encountering unreasonable sequence fragments or disordered regions; 2) certain structural fragments collapse when the quality of their corresponding generated tokens is relatively low.

\subsection{Analysis of Sampling Hyperparameters}
\label{asec:temp_cfg}

This section presents ablation studies on critical sampling hyperparameters across all tasks. 
We find that optimal sampling strategies vary with task characteristics.
A task with weak conditioning and a large solution space benefits from higher temperatures and more generation steps. In contrast, tasks with strong conditioning and narrow solution spaces perform better with lower temperatures, and sometimes even fewer sampling iterations.
Notably, our current implementation of classifier-free guidance (CFG) is only applicable to unconditional sequence-structure co-generation. 
In motif-scaffolding, CFG is mismatched with the implicit assumption that the structure track is conditioned on the sequence track. For protein structure prediction, the solution space is sufficiently small that external guidance can be counterproductive. 

\subsubsection{Unconditional Sequence-Structure Co-Generation}
\label{asec:temp_cfg_uncon_cogen}

\begin{figure}[h]
    \centering
    \includegraphics[width=\linewidth]{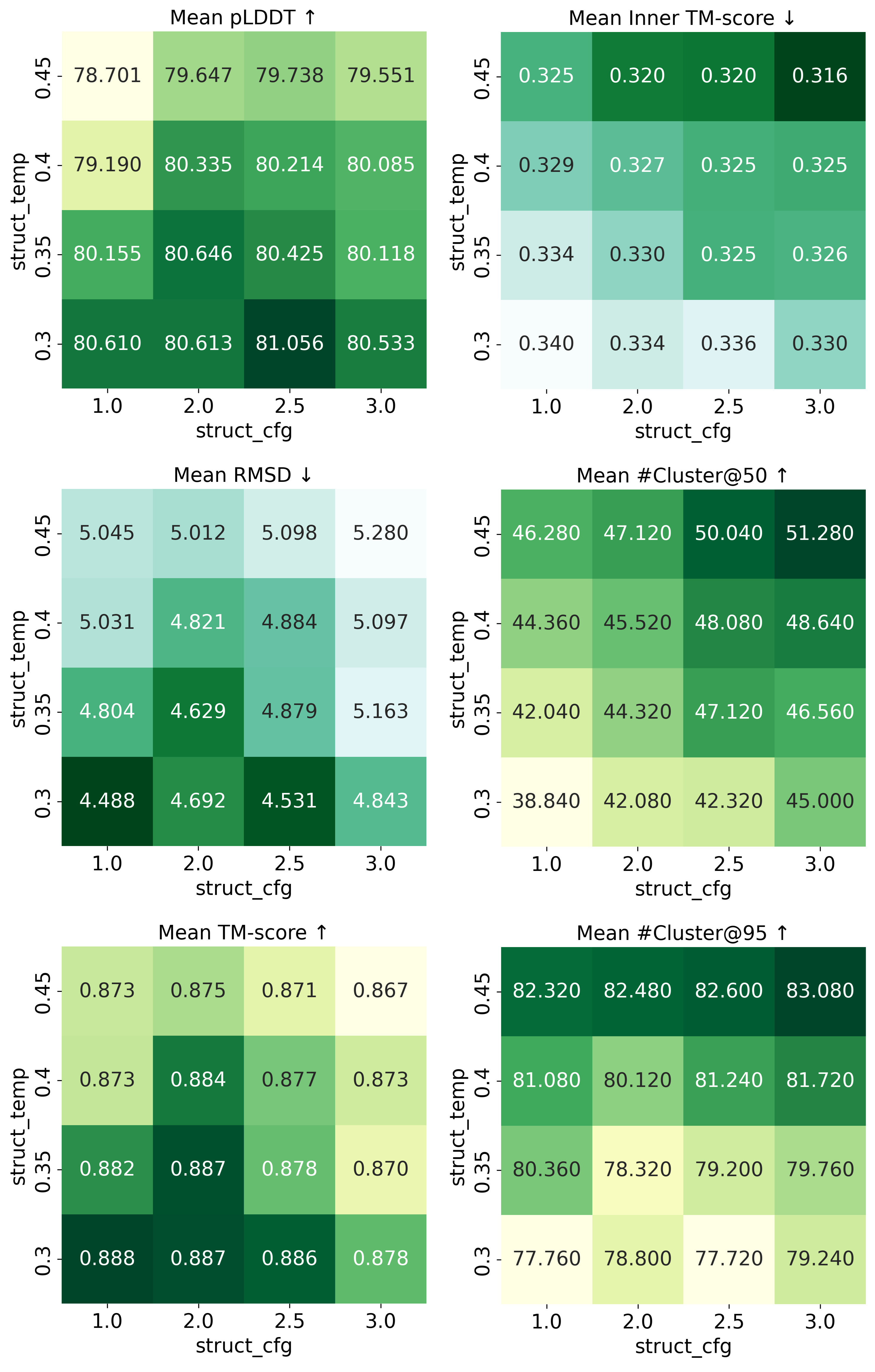}
    \captionsetup{font=small,skip=6pt}
    \caption{Unconditional Protein Sequence-Structure Co-Generation Performance of \ourname{} (155M)}
    \label{afig:hdprot_155_heatmap}
\end{figure}

\begin{figure}[h]
    \centering
    \includegraphics[width=\linewidth]{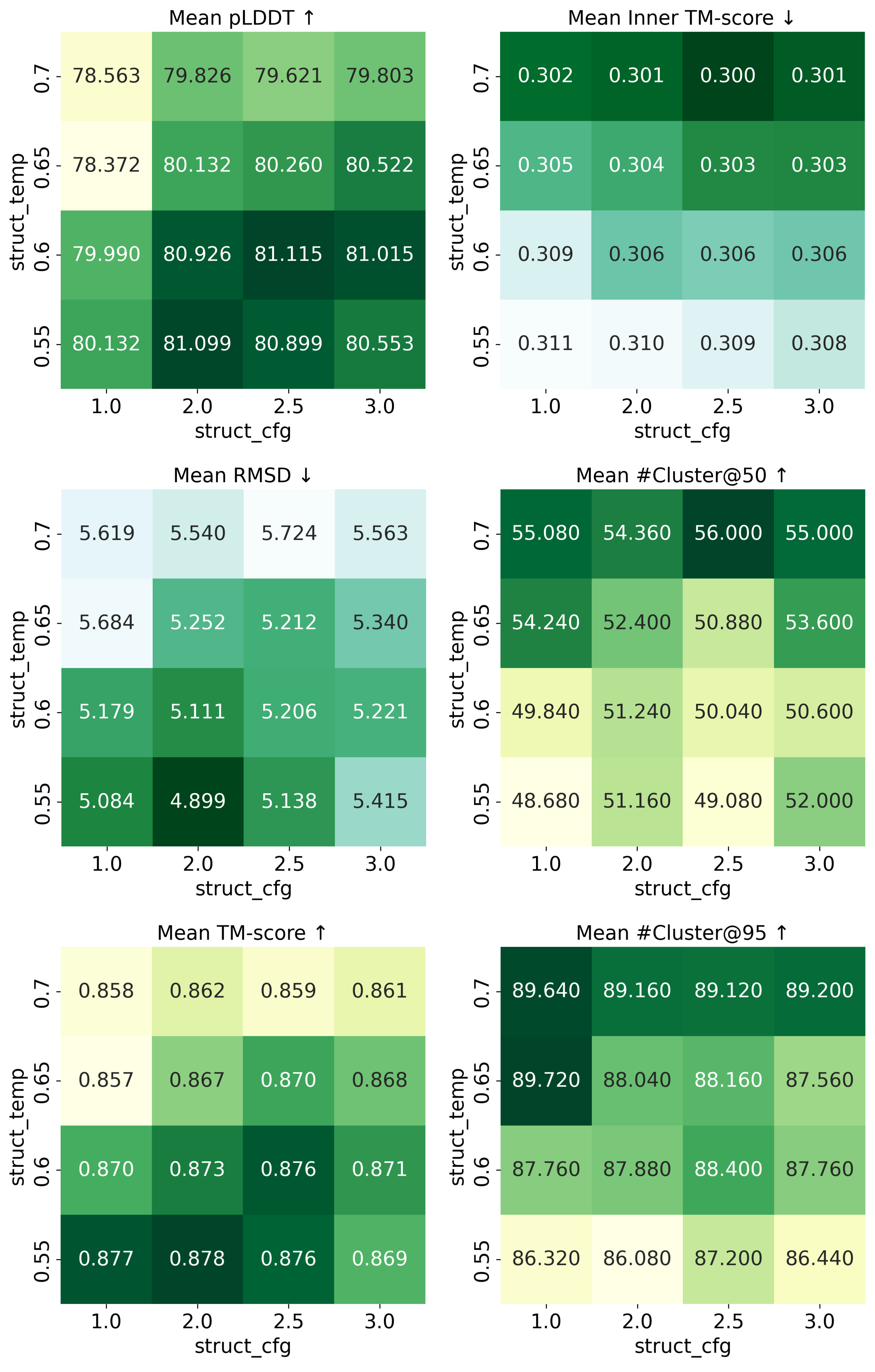}
    \captionsetup{font=small,skip=6pt}
    \caption{Unconditional Protein Sequence-Structure Co-Generation Performance of \ourname{} (670M)}
    \label{afig:hdprot_670_heatmap}
\end{figure}

For unconditional sequence-structure co-generation, the trade-off between the self-consistency and diversity in \ourname{}'s outputs is primarily governed by two hyperparameters: the structure sampling temperature $ \tau_z $ and the CFG scale.
We conduct a comprehensive grid search over four temperature values and four CFG scales for both \ourname{} (155M) and \ourname{} (670M), evaluating each configuration with five random seeds.
We report the average performance across the following metrics: pLDDT, scRMSD, scTM, Inner-TM\footnote{Inner-TM is the average pairwise TM-score among proteins of the same length.}, \#Cluster@50, and \#Cluster@95. 
Figure~\ref{afig:hdprot_155_heatmap} and Figure~\ref{afig:hdprot_670_heatmap} summarize these results as heatmaps, where darker colors indicate better performance.

Consistent with observations in other generative models, higher sampling temperatures increase diversity at the potential cost of sample quality. Empirically, we identify that \ourname{} (155M), a fully fine-tuned model, performs best with a lower temperature $\tau_z = 0.35$. In contrast, \ourname{} (670M) is a LoRA-tuned model where the modality expansion is somehow constrained, and a slightly higher temperature $\tau_z = 0.55$ is more beneficial.

Besides, moderate CFG strength, specifically with a CFG scale of $2.0$ or $2.5$, yields better pLDDT, scRMSD, and scTM scores than both no guidance (CFG scale $= 1.0$) and excessive guidance (CFG scale $= 3.0$).
At the same time, higher CFG scales generally improve diversity.
A CFG scale $ = 2.0 $ thus strikes a favorable balance between self-consistency and diversity in the generated proteins.

\subsubsection{Motif-Scaffolding}
\label{asec:temp_cfg_scaffolding}

Targeting a motif with a length of $l$ and a scaffold with a length of $L$, \ourname{} by default samples over $L-l$ steps to ``complete" the tokens other than the initial motif tokens. The structure sampling temperature is set as $ \tau_z = 0.1 $, and the classifier-free guidance is not introduced, which means the CFG scale $=1.0$. 
To demonstrate the optimality of these settings, Table~\ref{atab:motif_ablations} presents an ablation study by exploring three questions: 
\begin{itemize}
\item[1.] 
Should the structure sampling temperature be set at a higher level as in the unconditional co-generation, or should it be kept relatively lower ($\tau_z = 0.35/0.55 \text{ } vs. \text{ } 0.1$)? 
\item[2.] 
Is the classifier-free guidance effective in motif-scaffolding ($\text{CFG scale}=1.0 \text{ } vs. \text{ } 2.0$)? 
\item[3.] 
Should the initialized motif tokens be preserved, or should sampling be performed over all $L$ steps to overwrite them (Maintain Init. Motif Tokens $=$ True $vs.$ False)? 
\end{itemize}

\begin{table}[h]\footnotesize
    \centering
    \captionsetup{font=small,skip=3pt}
    \caption{Motif-Scaffolding Performance pf \ourname{}}
    \begin{tabular}{cccccc}
    \toprule
        \multirow{2}{*}{\makecell[c]{Model}} & \multirow{2}{*}{\makecell[c]{Struct.\\Temp.}} & \multirow{2}{*}{\makecell[c]{CFG\\Scale}} & \multirow{2}{*}{\makecell[c]{Maintain Init.\\Motif Tokens}} & \multirow{2}{*}{\makecell[c]{\#Solved / 24}} & \multirow{2}{*}{\makecell[c]{Success Rate}} \\
        \\
        \midrule
        \multirow{6}{*}{\makecell[c]{\ourname{}\\ (155M)}} & 0.1 & 1.0 & True & \textbf{18.2} (18, \textbf{19}) & \textbf{15.9\%} ± 0.3\% \\
        ~ & 0.1 & 2.0 & True & 17.8 (17, 18) & 15.1\% ± 0.9\% \\
        ~ & 0.1 & 1.0 & False & \textbf{18.2} (18, \textbf{19}) & 15.8\% ± 0.3\% \\
        \cmidrule(r){2-6}
        ~ & 0.35 & 1.0 & True & 18 & 15.1\% ± 0.7\% \\ 
        ~ & 0.35 & 2.0 & True & 17.6 (17, 18) & 14.2\% ± 0.6\% \\
        ~ & 0.35 & 1.0 & False & 18 & 15.1\% ± 0.7\% \\ 
        \midrule
        \multirow{6}{*}{\makecell[c]{\ourname{}\\ (670M)}} & 0.1 & 1.0 & True & \textbf{19.4} (19, \textbf{21}) & \textbf{24.1\%} ± 1.1\% \\
        & 0.1 & 2.0 & True & 18.2 (18, 19) & 23.2\% ± 0.5\% \\
        & 0.1 & 1.0 & False & 18.8 (18, 19) & 23.6\% ± 0.9\% \\
        \cmidrule(r){2-6}
        & 0.55 & 1.0 & True & 18.8 (18, 19) & 21.4\% ± 0.6\%  \\
        & 0.55 & 2.0 & True & 18.2 (18, 19) & 20.9\% ± 0.4\% \\
        & 0.55 & 1.0 & False & 18.6 (18, 19) & 21.5\% ± 0.6\% \\
        \bottomrule
    \end{tabular}
    \label{atab:motif_ablations}
\end{table}

The answers can be drawn from the experimental results. 
First, using a lower sampling temperature can slightly increase the number of solved problems and the average success rate. 
Second, introducing classifier-free guidance consistently degrades performance in motif-scaffolding. 
We attribute this to a mismatch between our current CFG implementation and the task objective. 
As introduced in the Appendix~\ref{asec:cfg}, CFG steers the generation of structure tokens toward greater alignment with the sequence track at each step. However, motif-scaffolding requires both the final sequence and structure tracks to align with the input motif, not merely with each other. 
Third, compared to sampling $L$ steps to overwrite the original motif tokens, sampling $L-l$ steps and preserving the initial motif tokens shows a slight advantage.

\subsubsection{Protein Structure Prediction}
\label{asec:temp_cfg_folding}

Protein structure prediction is typically regarded as a near one-to-one mapping task. From the perspective of conditional generation, the input sequence acts as a highly restrictive condition, leaving only a narrow structural solution space.
As shown in Table~\ref{atab:folding_ablations}, which compares various sampling strategies, the optimal approach in this low-entropy regime is to set the structure sampling temperature to $0.0$ and perform generation in a single deterministic step.
In contrast, increasing the sampling temperature, adding more iterative steps, or applying classifier-free guidance introduces unnecessary stochasticity, which degrades the accuracy of structure prediction rather than improving it.

\begin{table}[h!]\footnotesize
    \centering
    \captionsetup{font=small,skip=3pt}
    \caption{Protein Structure Prediction Performance of \ourname{}}
    \setlength{\tabcolsep}{2.2pt}
    \begin{tabular}{cccccccc}
    \toprule
        ~ & \multicolumn{3}{c}{Settings} & \multicolumn{2}{c}{CAMEO} & \multicolumn{2}{c}{PDB Date Split} \\
        \cmidrule(l){2-4} \cmidrule(l){5-6} \cmidrule(l){7-8}
        Model & Temp. & CFG & T & RMSD & TM-score & RMSD & TM-score \\
        \midrule
        \multirow{6}{*}{\makecell[c]{\ourname{}\\ (155M)}} 
          & $0.0$ & 1.0 & 1 & \textbf{9.199} ± 6.335 & \textbf{0.720} ± 0.200 & \textbf{6.231} ± 5.395 & \textbf{0.781} ± 0.181 \\
        ~ & $0.0$ & 1.0 & L & 9.699 ± 6.621 & 0.713 ± 0.200 & 6.654 ± 5.685 & 0.776 ± 0.185 \\
        ~ & $0.1$ & 1.0 & L & 9.716 ± 6.687 & 0.713 ± 0.200 & 6.653 ± 5.696 & 0.774 ± 0.188 \\
        ~ & $0.1$ & 2.0 & L & 9.607 ± 6.501 & 0.711 ± 0.200 & 6.599 ± 5.663 & 0.772 ± 0.187 \\
        ~ & $0.35$ & 1.0 & L & 9.734 ± 6.640 & 0.711 ± 0.200 & 6.648 ± 5.651 & 0.772 ± 0.187 \\
        ~ & $0.35$ & 2.0 & L & 9.637 ± 6.448 & 0.709 ± 0.199 & 6.592 ± 5.581 & 0.770 ± 0.188 \\
        \midrule
        \multirow{6}{*}{\makecell[c]{\ourname{}\\ (670M)}} 
          & $0.0$ & 1.0 & 1 & \textbf{7.468} ± 6.004 & 0.769 ± 0.177 & \textbf{5.001} ± 4.565 & 0.827 ± 0.153 \\
        ~ & $0.0$ & 1.0 & L & 7.500 ± 5.982 & 0.776 ± 0.177 & 5.023 ± 4.780 & \textbf{0.832} ± 0.149 \\
        ~ & $0.1$ & 1.0 & L & 7.525 ± 6.136 & \textbf{0.776} ± 0.176 & 5.014 ± 4.773 & 0.832 ± 0.150 \\
        ~ & $0.1$ & 2.0 & L & 7.743 ± 6.181 & 0.767 ± 0.177 & 5.084 ± 4.711 & 0.825 ± 0.150 \\
        ~ & $0.55$ & 1.0 & L & 7.757 ± 6.167 & 0.766 ± 0.177 & 5.065 ± 4.670 & 0.826 ± 0.150 \\
        ~ & $0.55$ & 2.0 & L & 7.848 ± 6.451 & 0.759 ± 0.178 & 5.131 ± 4.700 & 0.820 ± 0.152 \\
        \bottomrule
    \end{tabular}
    \label{atab:folding_ablations}
\end{table}

\subsubsection{Inverse Folding}
\label{asec:temp_cfg_inv_folding}

Inverse folding is typically regarded as a one-to-many prediction task. 
This task requires the generated sequence to adhere to the conditional structure, while allowing the exploration of diverse alternatives.
Table~\ref{atab:inv_folding_ablations} shows the comparison of the strategies for decoding each sequence token. It is observed that setting a small sampling temperature $\tau_z=0.1$ is better than setting a larger temperature $\tau_z=1.0$, and it is also better than directly using the deterministic argmax. That is to say, retaining few randomness is better than allowing excessive randomness, and it is also better than having no randomness at all. 

\begin{table}[h]\footnotesize
    \centering
    \captionsetup{font=small,skip=3pt}
    \caption{Inverse Folding Performance of \ourname{}}
    \setlength{\tabcolsep}{2.2pt}
    \begin{tabular}{ccccccc}
    \toprule
        ~ & \multicolumn{2}{c}{Settings} & \multicolumn{2}{c}{CAMEO} & \multicolumn{2}{c}{PDB Date Split} \\ 
        \cmidrule(l){2-3} \cmidrule(l){4-5} \cmidrule(l){6-7}
        Model & Strategy & Temp. & scRMSD & scTM & scRMSD & scTM \\ 
        \midrule
        \multirow{3}{*}{\makecell[c]{\ourname{}\\ (155M)}} & Vanilla & $0.1$ & \textbf{4.637} ± 4.730 & \textbf{0.863} ± 0.156 & 2.903 ± 3.683 & 0.919 ± 0.107 \\
        ~ & Vanilla & $1.0$ & 4.689 ± 4.812 & 0.862 ± 0.150 & 2.928 ± 3.694 & 0.919 ± 0.106 \\
        ~ & Argmax & - & 4.830 ± 4.935 & 0.861 ± 0.151 & \textbf{2.872} ± 3.511 & \textbf{0.919} ± 0.104 \\
        \midrule
        \multirow{3}{*}{\makecell[c]{\ourname{}\\ (670M)}} & Vanilla & $0.1$ & \textbf{4.675} ± 4.930 & \textbf{0.866} ± 0.151 & \textbf{2.871} ± 3.599 & \textbf{0.920} ± 0.103 \\ 
        ~ & Vanilla & $1.0$ & 4.750 ± 5.350 & 0.861 ± 0.152 & 2.944 ± 3.645 & 0.918 ± 0.103 \\ 
        ~ & Argmax & - & 4.708 ± 4.930 & 0.864 ± 0.146 & 2.900 ± 3.591 & 0.920 ± 0.103 \\ 
        \bottomrule
    \end{tabular}
    \label{atab:inv_folding_ablations}
\end{table}